# Non-contact acoustic micro-tapping optical coherence elastography for quantification of corneal anisotropic elasticity: in vivo rabbit study

Mitchell A. Kirby[1], Gabriel Regnault[1*], Ivan Pelivanov[1], Matthew O'Donnell[1], Ruikang K. Wang[1,3], Tueng T. Shen[2,3]

[1]Department of Bioengineering, University of Washington, Seattle, Washington 98105, USA

[2]School of Medicine, University of Washington, Seattle, Washington 98195, USA

[3]Department of Ophthalmology, University of Washington, Seattle, Washington 98104, USA

[*]*Correspondence: Gabriel Regnault, Department of Bioengineering, University of Washington, Seattle, Washington 98105, USA Email: gregnaul@uw.edu*

**Word count**

6833 words

**Funding Information**

This work was supported, in part, by NIH grants R01-EY026532, R01EY028753, R01-EB016034, R01-CA170734, and R01-AR077560, Life Sciences Discovery Fund 3292512, the Coulter Translational Research Partnership Program, an unrestricted grant from the Research to Prevent Blindness, Inc., New York, New York, and the Department of Bioengineering at the University of Washington.

**Commercial Relationships Disclosure**

Disclosure: **M. A. Kirby**, None; **G. Regnault**, None; **I. Pelivanov**, None; **M. O'Donnell**, None; **R. K. Wang**, None; **T. T. Shen**, None.

**Abstract (250 word limit)**

**Purpose.** To demonstrate accurate measurement of corneal elastic moduli in vivo with non-contact and non-invasive optical coherence elastography.

**Methods.** Elastic properties (in-plane Young's modulus $E$ and both in-plane, $\mu$, and out-of-plane, $G$, shear moduli) of rabbit cornea were quantified in vivo using non-contact dynamic Acoustic micro-Tapping Optical Coherence Elastography (A$\mu$T-OCE). The IOP-dependence of measured mechanical properties was explored in extracted whole globes following in vivo measurement. A nearly-incompressible transverse isotropic (NITI) model was used to reconstruct moduli from A$\mu$T-OCE data. Independently, cornea elastic moduli were also measured ex vivo with traditional, destructive mechanical tests (tensile extensometry and shear rheometry).

**Results.** Our study demonstrates strong anisotropy of corneal elasticity in rabbits. The in-plane Young's modulus, computed as $E = 3\mu$, was in the range of 20-44 MPa, whereas the out-of-plane shear modulus was in the range of 34-261 kPa. Both pressure-dependent ex vivo OCE and destructive mechanical tests performed on the same samples within an hour of euthanasia strongly support the results of A$\mu$T-OCE measurements.

**Conclusions.** Non-contact A$\mu$T-OCE can non-invasively quantify cornea anisotropic elastic properties in vivo.

**Translational Relevance.** As OCT is broadly accepted in Ophthalmology, these results suggest the potential for rapid translation of A$\mu$T-OCE into clinical practice. In addition, A$\mu$T-OCE can likely improve diagnostic criteria of ectatic corneal diseases, leading to early diagnosis, reduced complications, customized surgical treatment, and personalized biomechanical models of the eye.

**Introduction**

Early detection and monitoring of corneal diseases require better understanding of biomechanics to predict and prevent future corneal deformation. However, there are no current clinical tools that can accurately quantify corneal stiffness parameters that, along with the intraocular pressure (IOP), control corneal shape and focusing power. Personalized biomechanical models of the cornea can potentially be used to study disease progression and may play an important role in developing patient specific treatment plans. Non-invasive and non-destructive quantitative measurement of corneal mechanical moduli is needed so that personalized biomechanical models can be developed.

The macro-structure of corneal lamellae produce a highly anisotropic biomechanical deformation response depending on the type of applied force. As such, measured corneal stiffness is very different when a shear stress is applied along the lamellar plane[1–3] compared to a tensile stress along the same plane[4–7]. To account for the anisotropic deformation response, a nearly incompressible transverse isotropic (NITI) material model based on collagen fiber macro-structure was developed. It was shown that the cornea must be described by at least two shear moduli (in-plane tensile and out-of-plane shear moduli) [8], a departure from commonly used single-modulus models. Additionally, it was shown that elastic waves in dynamic OCE studies of the cornea were accurately described by the NITI model. Non-contact acoustic micro-tapping optical coherence elastography (AµT-OCE) was used to measure anisotropic mechanical properties in ex vivo porcine cornea, where the cornea's bounded structure forced guided wave propagation and enabled reconstruction of in-plane, $\mu$ (where $E = 3\mu$), and out-of-plane, $G$, shear moduli by fitting experimentally obtained dispersion curves in the wavenumber-frequency domain to the theoretical (NITI) model.[8]

The NITI model was further explored using both AµT-OCE and traditional destructive mechanical testing.[7] Specifically, the Young's modulus, $E$, and out-of-plane shear modulus, $G$, of fresh porcine whole globes inflated to controlled intraocular pressures (IOP) were measured with AµT-OCE before being cut into corneal buttons and tested using a parallel-plate rheometer and then stripped and tested via in-plane extension loading. Results were consistent with an order of magnitude difference between in-plane, $E$, and out-of-plane, $G$, elastic moduli for all testing methods. While there were some differences in corneal condition during testing (such as curvature, boundary conditions, loading type, preconditioning, etc.), it was shown that moduli quantified from OCE data analyzed with the NITI model were accurate and provided a non-contact, non-destructive path to measure corneal anisotropic biomechanical properties.

Because the host-imaging method in AµT-OCE (Optical Coherence Tomography, OCT) can accurately (and without contact) image corneal structure and shape in addition to quantifying mechanical properties, AµT-OCE can potentially provide the in vivo measurements required for personalized biomechanical models of the cornea. However, robust and reliable measurement of anisotropic elastic moduli has not yet been demonstrated in living species. Although elastic properties of in vivo rabbit cornea were estimated previously with OCE, the assumed material models resulted in inaccurate values of elastic moduli because they did not account for cornea bounding and/or cornea mechanical anisotropy.[9–11] Thus, to date, there are no studies demonstrating non-contact, non-invasive measurements of cornea shear elastic anisotropy in vivo.

The primary goal of this study was to test the AµT-OCE method using an in vivo rabbit model (New Zealand White Rabbits). It was chosen due to the similarities in geometrical and functional properties with human cornea. In the present study, rabbit cornea elastic moduli were measured with AµT-OCE in vivo and the intraocular pressure (IOP) was recorded. Following in vivo measurements, whole globe samples were extracted and inflated to controlled IOP for comparison between in vivo moduli and ex vivo, pressure-dependent values. Finally, samples were sectioned and tested using both a parallel-plate rheometer (for independent determination of out-of-plane shear modulus, $G$) and tensile extensometer (for independent determination of in-plane Young's modulus, $E$). The results demonstrated that AµT-OCE can reliably generate and track elastic waves in rabbit cornea in vivo, enabling accurate reconstruction of corneal elasticity. In vivo results were compared with both controlled ex vivo and destructive mechanical tests that currently serve as the gold standard in biomechanical testing.

**Methods & Materials**

*Sample preparation*

In this study, five (5) adult New Zealand White Rabbits (3 Female, 2 Male, mean ~4 kg) were acquired and housed at the University of Washington Vivarium for a minimum of 5 days prior to experiments. All rabbits were treated in accordance with the Association for Research in Vision and Ophthalmology Statement for the Use of Animals in Ophthalmic and Vision Research. All procedures were approved by UW IACUC (PROTO202000139). While in vivo tests with rabbits do not exactly match the blinking, respiration, and saccades seen in humans, this animal study presents a good model for similar types of motion expected in human cornea for clinical scans.

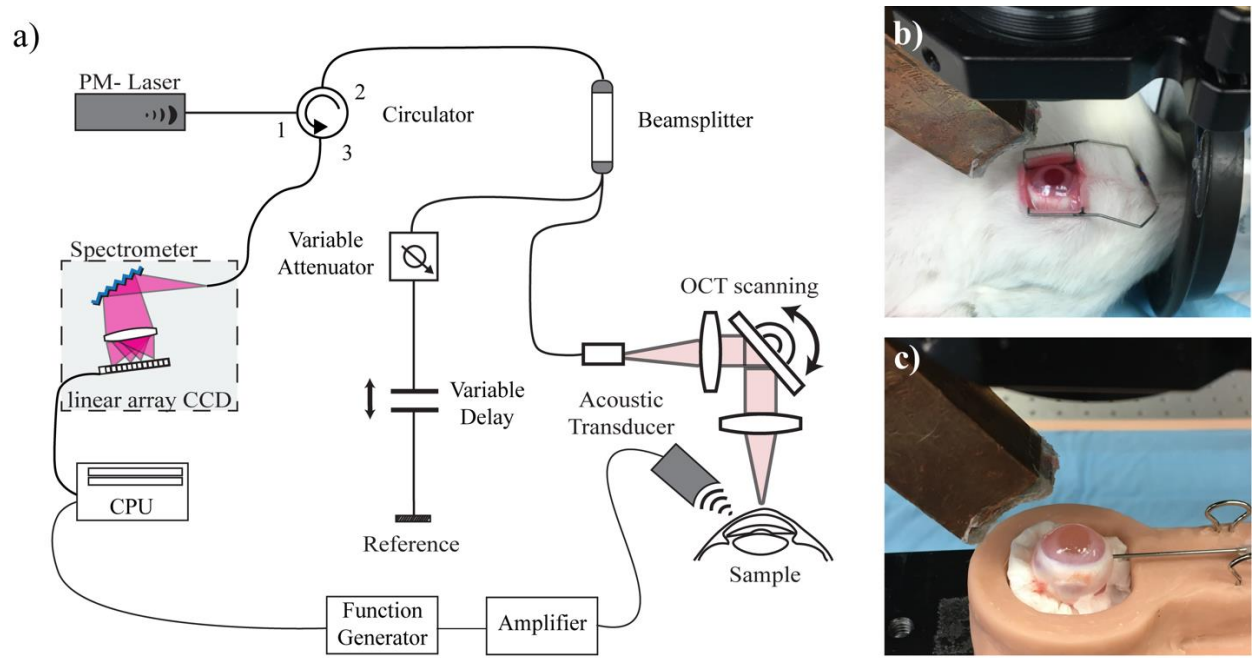

**Figure 1.** a) Schematic of AµT-OCE system b) in vivo imaging of rabbit cornea c) ex vivo whole globe with needle insertion to control IOP.

Each rabbit was transported from the UW vivarium to a research facility for imaging with veterinarian assistance. A trained veterinarian staff anesthetized the rabbit using a one-time dose of 2 mg/kg Xylazine and 50 mg/kg Ketamine, followed by a secondary dose of 15 mg/kg ketamine at the mid-point of the experiment. All animals were kept on 1-5% isoflurane until in vivo testing was completed.

For in vivo OCE testing, rabbits were placed on their side on a sterile pad under the imaging arm of the OCE system (Figure 1a). Sterile cotton pads were used to stabilize the rabbit's head and align the eye for imaging. The eyelid was held open using a pediatric eyelid speculum (Figure 1b) and the intraocular pressure of each eye was tested using a contact tonometer (Tono-Pen XL, Reichert, Inc., Buffalo, NY). Following tonometry, at least 10 non-contact OCE scans were performed on each eye.

Following imaging, rabbits were euthanized using a 1.5 ml Euthasol injection and whole globe corneas were harvested no more than 30 minutes following initial anesthesia. All excess tissue was removed to expose the sclera, and each globe was rinsed with balanced saline solution (BSS).

Whole globes were placed in a mold containing a damp sterile cotton pad to stabilize samples and mimic in vivo boundary conditions (Figure 1c). A 20-gauge needle connected to a bath filled with BSS was inserted through the temporal wall of the sclera to apply a controlled internal hydrostatic pressure (intraocular pressure, IOP). The IOP was controlled by raising and lowering the bath to impart controlled pressures between 1 and 18 mmHg, at increments of 2 mmHg. Each sample was held at the corresponding pressure for 5 minutes before scanning, over which BSS was applied lightly to prevent corneal dehydration. Five (5) repeat OCE scans were acquired at each pressure. Each sample was scanned at room temperature and imaging took no longer than 1 hour per sample.

Following OCE, cornea-scleral rings were extracted for mechanical testing, where each ring was used to test the out-of-plane shear modulus, $G$, via rotational rheometry.

Following rheometry, each sample was cut into strips for tensile testing of the in-plane Young's modulus, $E$, using an extensometer. All tests were performed within 6 hours of euthanasia and all samples were transported using cloths dampened with BSS.

### *Nearly incompressible transverse isotropic (NITI) model*

The macro-structure of corneal lamellae creates a highly anisotropic deformation response to force. The primary lamellae orientation (arranged along in-plane layers[12,13]) contributes to a stress-strain relationship that can be approximated using four independent elastic constants $\lambda$, $\delta$, $G$, and $\mu$ (assuming small-strain deformation).[7,14] The elasticity matrix in the cornea (in Voigt notation) can thus be described as:

$$\begin{bmatrix} \sigma_{xx} \\ \sigma_{yy} \\ \sigma_{zz} \\ \tau_{yz} \\ \tau_{xz} \\ \tau_{xy} \end{bmatrix} = \begin{bmatrix} \lambda + 2\mu & \lambda & \lambda & & & \\ \lambda & \lambda + 2\mu & \lambda & & & \\ \lambda & \lambda & \lambda + \delta & & & \\ & & & G & & \\ & & & & G & \\ & & & & & \mu \end{bmatrix} \begin{bmatrix} \varepsilon_{xx} \\ \varepsilon_{yy} \\ \varepsilon_{zz} \\ \gamma_{yz} \\ \gamma_{xz} \\ \gamma_{xy} \end{bmatrix} \qquad (1)$$

where $\sigma_{ij}$ denotes engineering stress, $\varepsilon_{ij}$ denotes engineering strain, $\tau_{ij}$ denotes shear stress, $\gamma_{ij} = 2\varepsilon_{ij}$ denotes shear strain, and the subscripts $x$, $y$, and $z$ denote standard Cartesian coordinates. Because the cornea is nearly incompressible, the longitudinal modulus $\lambda$ does not define deformation, and the corneal strain response to an applied stress can be fully defined by the out-of-plane and in-plane shear modulus ($G$ and $\mu$, respectively), and an additional term, $\delta$.

As shown previously, the in-plane Young's modulus depends on both $\mu$ and $\delta$.[7] However, corneal structure constrains the latter to the range $-2\mu < \delta < 0$, which restricts the Young's modulus to the range:

$$2\mu < E_T < 3\mu. \qquad (2)$$

Although $\delta$ cannot be determined from guided wave propagation, the limited range for $E_T$ suggests that $\mu$ and $G$ can provide very close approximations of full cornea deformations assuming small deformation. While the exact relationship between the in-plane Young's modulus ($E_T$) and $\delta$ has not yet been determined, we assume tensile isotropy here so that $E = E_T \cong 3\mu$. The approximation of $\delta = 0$ assumes corneal tensile isotropy and likely leads to a slight overestimation of $E_T$.

*Acoustic micro-tapping OCE system*

A home built AμT-OCE system launched mechanical waves in the cornea and tracked their propagation in space and time. The system in Figure 1a has been detailed previously.[7] Briefly, elastic waves were generated using a cylindrically focused air-coupled piezoelectric transducer (AμT) driven with a 100 μs-long chirped (1 MHz-1.1 MHz) waveform providing a temporally localized and spatially focused acoustic 'push'. The resulting elastic wave was measured using a stable Michelson-type fiber-optic interferometer where a broadband superluminescent diode (SLD1018P, Thorlabs, NJ) with central wavelength 1310 nm (45 nm full-width-half-maximum bandwidth) was coupled into polarization maintaining fibers and components for depth encoded (1.5 mm effective imaging range) OCT imaging and motion detection.

The generation and tracking of propagating elastic waves were achieved using an acoustic pulse aligned in time with the start of multiple (256 consecutive) OCT A-scans performed at the same spatial location (referred to as an M-scan). Sequential acoustic pulses were then generated at a fixed location while the OCT M-scan was performed at different locations (256 spatial locations) perpendicular to the AμT line source (generating what is referred to as an MB-scan[15]). Each MB-scan provided a 3D dataset with 1024 depth x 256 lateral x 256 temporal dimensions. The scan rate of the system (determined by the line-scan camera) was 90 kHz, corresponding with a total scan time of around 750 milliseconds. For ex vivo scans at low IOP, 512 temporal scans were taken to allow the elastic wave to propagate across the full 10 mm field of view.

The MB data-set was used to reconstruct elastic wave propagation based on local vertical particle vibration velocity,[16] where the measured displacement sensitivity of the system was approximately 1 nm in water. The log-compressed real-part of the OCT signal was used to reconstruct corneal structure, and the surface of the cornea was detected using an automatic segmentation algorithm. The vibration velocity along the surface of the cornea was determined using a weighted-average (one half of a Gaussian window, HWHM = 90 μm, weight decreasing with depth) along the anterior 183 μm of the cornea, providing raw space-time (*x-t*) maps of the vertical displacement from propagating guided elastic waves detected along the air-tissue boundary.

For in vivo imaging, bulk motion due to pulsatile vibrations, breath, and rabbit movements can make it difficult to accurately reconstruct elastic wave propagation. In this study, broadband elastic waves generated by the acoustic excitation travel with energy concentrated in the multiple-kHz range. Because much of the bulk motion associated with in vivo imaging occurs in the low-frequency range, a temporal bandpass filter was applied to *x-t* data where vibration frequencies below 50 Hz and above 4 kHz were removed. Additionally, randomly propagating diffuse wavefields[17] were limited by a directional filter, where a two-dimensional Fast Fourier Transform (2D FFT) was performed on the *x-t* plot and a mask applied to the 2nd and 4th quadrant of the *k-f* space, followed by an inverse 2D FFT.[18] Finally, unwanted reflections and forward propagating waveforms from sidelobes in the acoustic excitation, or remaining diffuse propagations, were removed by applying a moving temporal window centered on the peak of the vertical velocity in the *x-t* plot. The moving temporal window utilized a super-Gaussian ($SG$) function (Eq. 3) that followed the maximum vibration velocity of the wavefield $t_m^{wf}(x)$ at each discrete position $x$:

$$SG(t) = \exp\left[-\left(\frac{1}{2}\left(\frac{t-t_m^{wf}(x)}{\sigma_t}\right)^2\right)^2\right] \tag{3}$$

with $\sigma_t$ = 0.5 ms. The resulting surface *x-t* information was used to reconstruct elastic moduli assuming a NITI material.

### *Reconstruction of corneal elastic moduli*

Since the cornea is a bounded anisotropic material, an appropriate model is required to describe complex wavefields and perform accurate analysis for moduli reconstruction. As shown previously and described above, a bounded nearly incompressible transversally isotropic (NITI) material model can be used for elastic wave propagation in the cornea.

Once surface waveforms were captured (Figure 2a), the space-time (*x-t*) plots (Figure 2b) were subject to a 2D FFT to display the waveform in wavenumber-frequency (*k-f*) space. An inversion method based on the solution to guided elastic waves in a NITI material quantified out-of-plane shear modulus, $G$, as well as in-plane Young's modulus $E$ (assuming $E_L=E_T$).[7,8] As the resolution limit (determined in guided materials by the wavelength and corneal thickness[19]) was approximately 1 mm, the corneal surface area available for imaging (~10 mm) was sufficient for accurate reconstruction of elastic moduli.

Because only the $A_0$ -mode is expected in the cornea for the frequency ranges produced by AµT, the $A_0$ -mode solution (detailed previously[8]) was solved in *k-f* space for a broad range of input shear moduli ($G$, and $\mu$) with a fixed corneal thickness. The central cornea thickness was measured in each scan using the OCT image assuming a refractive index of 1.38[31].

An iterative routine estimated in-plane tensile and out-of-plane shear moduli ($\mu$ and $G$, respectively), where the best-fit theoretical dispersion relation was performed by maximizing the following objective function using simplex optimization (*fminsearch*, MATLAB, MathWorks, Natick, MA):

$$\Phi(\mu, G) = \frac{1}{N_f}\sum_f \sum_k w(f,k;\mu,G)|\hat{v}(f,k)|^2 - \beta \left|\frac{\mu}{\lambda}\right|. \qquad (4)$$

The normalized 2D Fourier spectrum in *k-f* space is $\hat{v}$, the $A_0$ mode solution for a NITI material is decribed by $w(f,k;\mu,G)$, and the nearly incompressible assumption is defined by $\beta$ (set to 1 based on an L-curve analysis).[2] The value of $\lambda$ was updated at each iteration according to $\lambda = \rho c_L^2 - 2\mu$. The corneal density was assumed to be $\rho$ = 1000 kg/m³ and corneal longitudinal wave speed $c_L$ = 1540 m/s. The cornea was assumed bounded from below by water with a density of 1000 kg/m³ and longitudinal wave speed of 1540 m/s. To avoid convergence to a local (as opposed to global) maximum in Eq. (4), five independent fits were performed, with quasi-random (within a reasonable range of expected values) initial values of $G_0$ and $\mu_0$. The final output of $\mu$ and $G$ were set to those corresponding to the highest value in Eq. 4. An example of the resulting best-fit $A_0$-mode superimposed on top of the *k-f* spectrum in a rabbit cornea is shown in Figure 2c.

While the optimization function provides the theoretical $A_0$ mode that most closely matches experimental data, it does not provide information on the fit quality. Although the best-fit line generally follows the spectral maxima (Figure 2c), the degree to which the theoretical $A_0$ mode follows the measured spectral maxima is not directly determined by the fitting regime. By normalizing the best-fit line to the unconstrained global maximum of the spectral peaks at each spectral bin in *k-f* space ($\Phi_{\max}$), the quality of the fit can be estimated:

$$g_{\text{NITI}} = \frac{\Phi_{\text{NITI}}}{\Phi_{\max}}, \tag{5}$$

where $g_{\text{NITI}}$ describes the fraction of the maximum mode energy captured by the best-fit A₀ dispersion curve. A value of $g_{\text{NITI}} = 1$ would indicate that the experimental dispersion curve has the exact match with the NITI model ($\Phi_{\text{NITI}}$ captures all the measured mode's maximum energy). Because low values of $g_{\text{NITI}}$ correspond with a failure to converge to a solution that captures most of the mode energy, it can be used to exclude unreliable scans (due to misalignment, rabbit motion, model inaccuracies, etc.). The output values of $G$ and $\mu$ were considered inaccurate for any fit that corresponded to $g_{\text{NITI}}$ below 0.87, and 0.92, respectively. Details on determining exclusion parameters can be found in Supplementary Methods.

Due to the maximization approach for weighted fitting, residual errors are not computed. To estimate uncertainty in the final fit value, $G$ and $\mu$ were varied independently around the optimum values of $\Phi_{\text{NITI}}$. For each combination of $G$ and $\mu$, Eq. 4 was used to calculate $\Phi(\mu, G)$, which was then used to determine:

$$\psi(\mu, G) = \frac{\Phi(\mu,G)}{\Phi_{\max}}. \tag{6}$$

Multiple fits were performed on independent scans of the same sample, and both model and system error were considered to determine uncertainty intervals around the best-fit values. The best-fit estimate ($g_{\text{NITI}}$) was determined independently for at least 5 repeat scans (in both in vivo and ex vivo data sets). The standard deviation of $g_{\text{NITI}}$ was then used to determine the cut-off value in $\psi$ for each set of scans.

Consider the representative example shown in Figure 2d, e, where the mean value of $g_{\text{NITI}}$ ($g_{\text{NITI}} = \psi_{\max}$) was 0.98 with a standard deviation of 0.003 (or 0.3%) across 5 independent scans. The variation in $g_{\text{NITI}}$ determined the uncertainty interval using the corresponding $G$ and $\mu$ values for $\psi$ at 0.3% below the max (i.e. $\psi = .977$ for $G$ and $\mu$). This method produced uneven error bars due to the shape of $\psi$. Note that the absolute value of $g_{\text{NITI}}$ provides an estimate of model error. When $g_{\text{NITI}}$ is reduced, the shape of $\psi$ widens for both moduli and, therefore, model uncertainty increases.

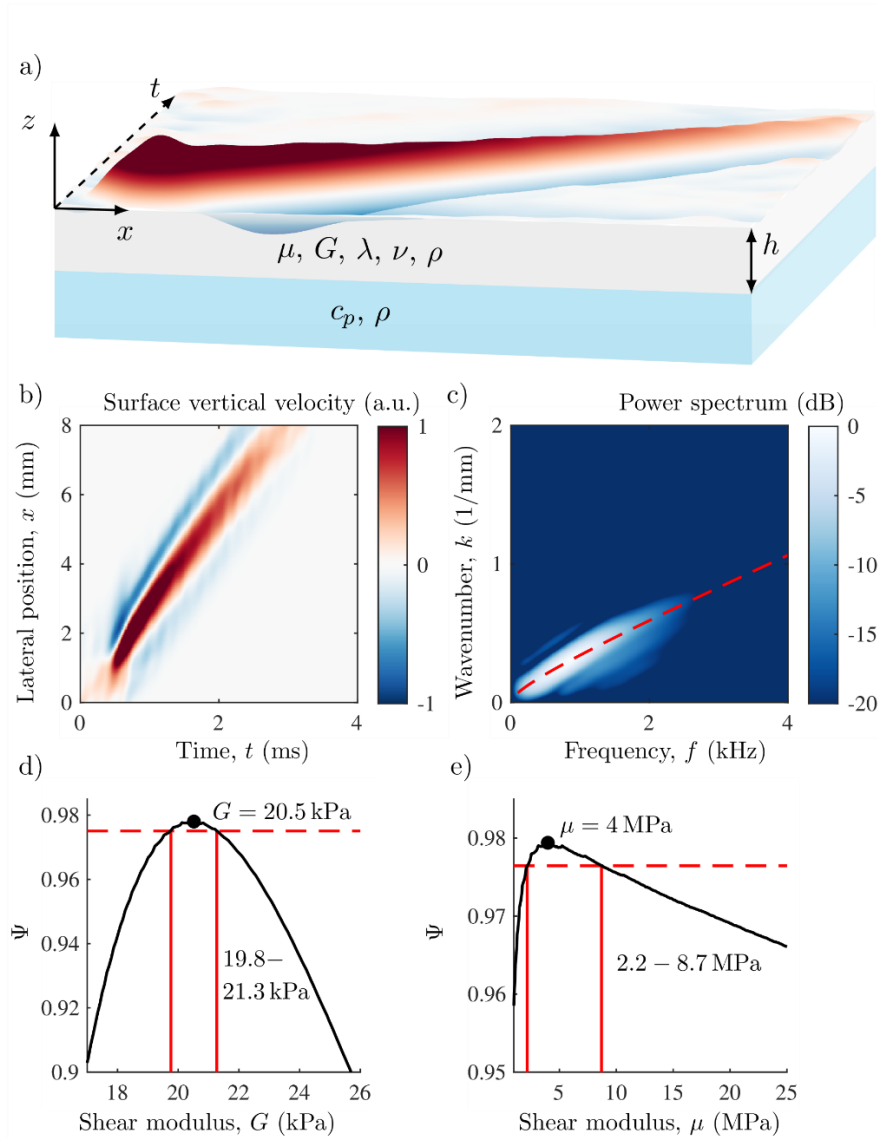

**Figure 2.** a) Measured corneal surface vibrations in a whole globe rabbit sample inflated to a pressure of 7 mmHg. b) Filtered and windowed *x-t* plot. c) Best fit of $A_0$ mode plotted on top of *k-f* plot. d) Optimization function showing best-fit and uncertainty around $G$ and e) $\mu$. Uncertainty intervals (red lines) calculated for the representative example where standard deviation in $g_{\text{NITI}}$ was 0.3%. The best-fit solution provided $G = 20.5$ kPa, with uncertainty of 19.8 kPa – 21.3 kPa, and $\mu = 4.0$ MPa, with uncertainty of 2.2 MPa – 8.7 MPa.

Because $n = 5$ repeat scans were taken, five independent measurements produced corresponding values for $G$ and $\mu$, and their respective uncertainty ranges. The mean value of $G$ and $\mu$ for all repeat fits was considered for each cornea (at each IOP). Uncertainty ranges were calculated as mean-squared values of lower and upper uncertainty limits in independent scans. The process provided a value for the OCE-measured out-of-plane shear modulus, $G$, and in-plane Young's modulus, $E$ (assuming $E = 3\mu$), as well as error bars associated with the uncertainty of reconstruction. Uncertainty intervals for each independent OCE measured value can be found in the Supplementary Material.

### *Mechanical testing of ex vivo samples*

Following OCE testing of corneas, corneo-scleral rings were extracted and used to test the out-of-plane shear modulus, $G$. A rheometer (Anton Paar MCR 301 Physica) assessed the frequency-dependent shear behavior (storage, $G'(\omega)$, and loss, $G''(\omega)$) of corneal buttons over a range of 0.16-16 Hz. A 5 N compressive preload was applied and the peak shear strain was ~.1%. The test was performed twice, and the mean of each run was used for the final value.

Following rheometry, corneal-sclera buttons were cut into strips and pneumatically clamped (2752-005 BioPuls submersible pneumatic grips, 250 N max load). A 50 mN pre-load was applied to each sample and stretched at 2 mm/min up to 10% strain. Two load-unload cycles were performed to precondition the tissue. Three rounds of force-elongation followed by relaxation were performed and converted to stress-strain according to sample geometry. A second order exponential was fit to three-sets of raw data to determine the stress-strain curve. The in-plane Young's modulus, $E$, was defined as the tangential slope of the stress-strain curve. Extension testing provided a value for the strain-dependent Young's modulus, $E$, up to 10% strain.

**Results**

*In vivo pressure (Tono-Pen)*

In vivo intraocular pressure was measured in each eye using a contact tonometer (Tono-Pen XL, Reichert, Depew, NY). Because contact tonometers have been reported to have significant inaccuracies (with errors increasing with greater IOP),[20,21] a correction factor was applied to all in-vivo tonometry measurements to facilitate comparison between in vivo and ex vivo measurements. The correction factor in this study was determined using a direct comparison of Tono-Pen measurements made on 5 ex vivo whole globe samples inflated to known intraocular pressures.

As the inflation pressure was incrementally raised from 3 mmHg to 21 mmHg in a subset of ex vivo whole globes, Tono-Pen measurements were recorded at each pressure (Figure 3). The dependence of the Tono-Pen -measured IOP on the inflated pressure was averaged over 5 independent measures and then fit with a linear function:

$$IOP_{Tonopen} = 0.3 * IOP_{Control} + 6 \ . \tag{7}$$

As such, in vivo IOP values underwent a correction factor associated with Eq. (7), where $IOP_{Control}$ (corresponding to a coarse estimate of the 'actual' IOP) was found for each Tono-Pen determined value ($IOP_{Tonopen}$). The result of this test suggests that a Tono-Pen measured value of 10 mmHg, for example, corresponds to an actual value of 13 mmHg (with uncertainty associated with the standard deviation of the values at each pressure between ~6 mmHg and 21 mmHg). The mean difference between the Tono-Pen and actual values was ~6.5 mmHg across all pressures for the sample size used. As such, error bars corresponding to ± 6.5 mmHg are included in the adjusted in vivo pressure values displayed in Figure 4. All error bars were cut-off at 0 mmHg for display.

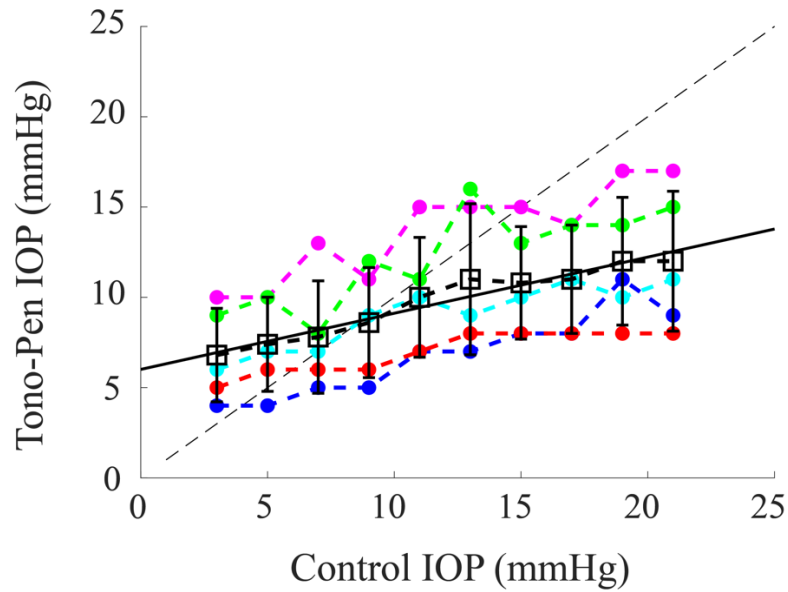

**Figure 3.** Tono-Pen measured IOP value (mmHg) as the internal pressure was raised using a lifted water bath. Colored dots and lines correspond with each sample, black squares and error bars denote mean and standard deviation, respectively, and the black solid line corresponds with the best-fit linear function (Eq. 7). The dotted black line is the one-to-one (slope=1) line for visualization.

*Quantification of elastic moduli*

Anisotropic elastic moduli were quantified for each sample in vivo, following extraction (ex vivo), and via destructive mechanical testing for each sample. A summary of the values measured via OCE is shown in Figure 4. In vivo measurements of $G$ (blue squares) and $E$ (red square**s**) are plotted at the corrected pressure (horizontal error bars are associated with the range of possible 'actual' pressure values). Ex vivo moduli (triangles) were measured and displayed at the actual IOP value. The vertical error bars in OCE measured values correspond with the standard deviation of the mean in each cluster, at each IOP. Note that of the 10 samples tested, one had a fit quality below the exclusion criteria and was omitted. Because the goodness of fit generally decreased with increasing IOP for both moduli, there are fewer data-points at pressures greater than 17 mmHg (Supplementary Methods).

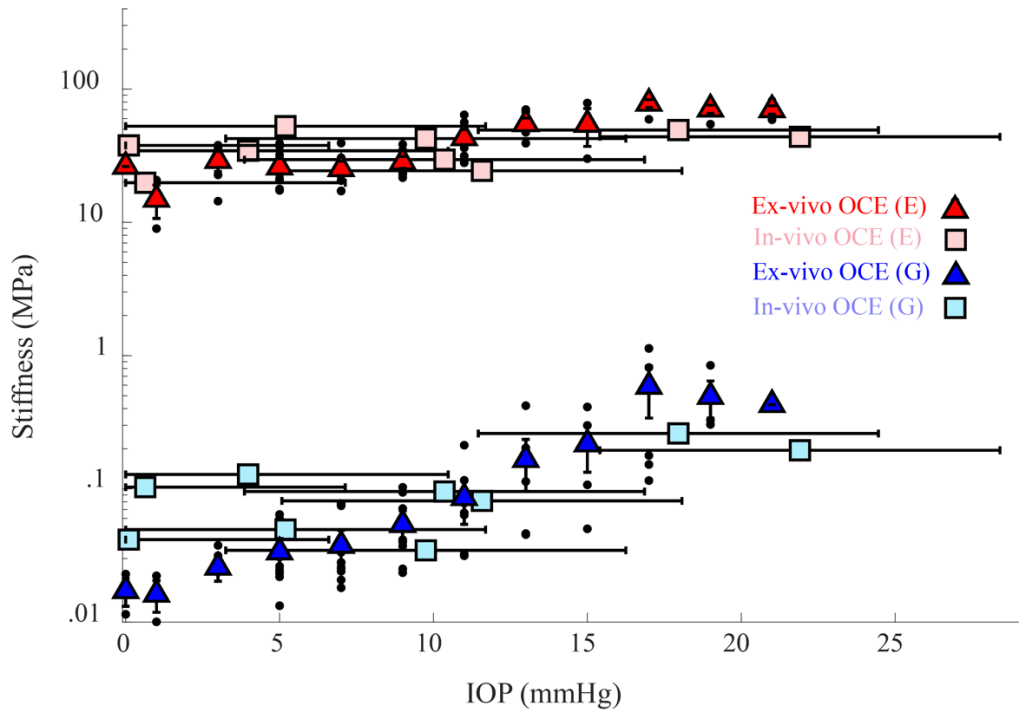

**Figure 4**. Summary of elasticity measurements performed in vivo and ex vivo. The squares denote in vivo values, and the triangles are the mean of ex vivo measurements. Each black dot is the OCE measured ex vivo modulus for a single sample at the associated pressure (all results shown in Supplementary Materials).

A summary of the anisotropic elastic moduli measured in vivo, ex vivo, and with destructive mechanical testing is presented in Figure 5. OCE- measured values represent the mean and standard deviation of the entire group of 9 samples. For in vivo samples, the IOP range (after correction using Fig.3) 11 mmHg on average. The range of in vivo $G$ was 34 kPa- 261 kPa, and for $E$ was 20 MPa- 44 MPa. For both in vivo and ex vivo whole globes measured via OCE, the stiffness generally increased with increasing pressure. In ex vivo whole globes, $G$ increased from 31 kPa ($\pm$ 15 kPa) in the IOP range from 3 mmHg to 5 mmHg, to 98 kPa ($\pm$65 kPa) in the IOP range from 11 mmHg to 13 mmHg, and $E$ increased for the same IOP ranges respectively from 27 MPa ($\pm$ 9 MPa) to 47 MPa ($\pm$ 13 MPa). The values are not reported at higher pressures due to decreased sample size with reliable quality of fit. The mean and standard deviation of rheometry measured values of $G'(\omega)$ (Figure 5a) at 16 Hz was 75 kPa ($\pm$ 43 kPa). Note that shear rheometry was performed over a lower range of frequencies than OCE, which presumably should lead to a slightly lower estimate of G, as shown. The tensile modulus, $E$, increased with strain from 2.8 MPa ($\pm$ 1.1 MPa) at 1% to 32 MPa ($\pm$ 20 MPa) at 10%. The mean and standard deviation of the high strain value is displayed in Figure 5c. Corresponding frequency-dependent values of $G'(\omega)$ and $G''(\omega)$ (measured in all corneal buttons with shear rheometry), as well as the strain-dependent value of the Young's Modulus, $E$, along with OCE measured values, can be found in Supplementary Methods for each sample included in the analysis. As can be seen in Figure 5, the range of values measured with all methods (for comparable IOP in OCE) were in close agreement.

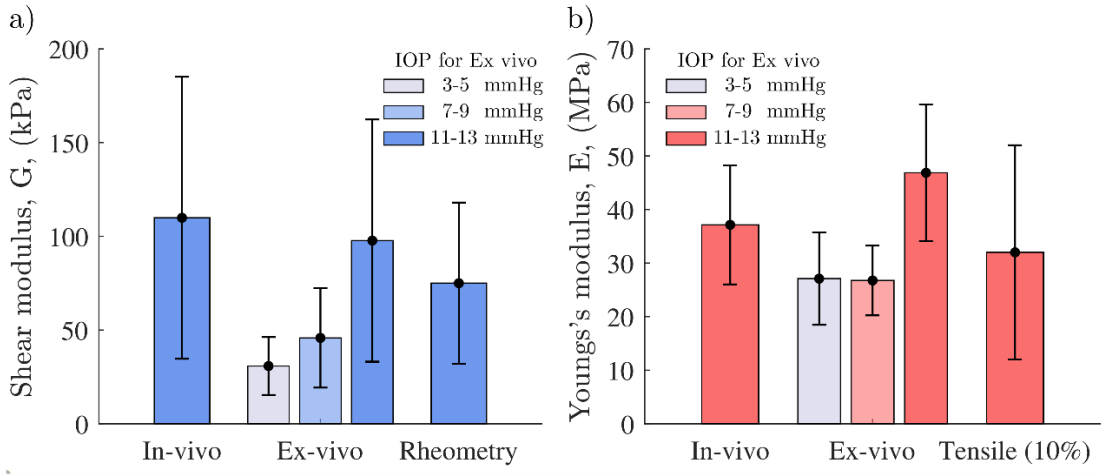

**Figure 5.** The mean and standard deviation for all corneas measured via OCE in vivo and ex vivo, as well as by destructive mechanical testing. a) mean shear, $G$, modulus and b) Young's, $E$, modulus for: in vivo OCE at the fixed physiological IOP (measured with the tonopen and equal to 11 mmHg on average after the correction presented in Fig.3), ex vivo OCE for three different ranges of IOP indicated in the legend, the mean rheometry value of the storage modulus at 16 Hz, and tensile testing value of the tangential modulus at 10% strain. Standard deviation corresponds to the deviation of measured moduli across the population of 9 cornea samples.

**Discussion**

The non-contact nature and consistency of AµT-OCE make it a promising method to evaluate corneal elasticity, monitor keratoconus, and potentially serve as a tool to develop personalized biomechanical models that can predict corneal response to ophthalmic interventions. Unlike other imaging methods, the host imaging modality of OCT allows fast, non-contact, potentially high-resolution imaging due to scanning-point focusing of the laser light used to recreate structural images and track propagating elastic waves.

AμT-OCE potentially can change diagnostic criteria of ectatic corneal diseases, leading to early diagnosis, reduced complications, customized surgical treatment, and new opportunities to develop personalized biomechanical models of the eye.

Optimizing the imaging instrument for human use requires live animal testing because there are no equivalent phantoms that mimic the low-frequency motion associated with both cardiac and respiratory cycles. Because the animals used in this study were anesthetized and stable, bulk motion associated with skeletal muscular movement was minimal over the ~3 second scan time. Singh *et al.*[32] have shown (in their case for anesthetized rodents) that the frequency of respiratory motion and the heartbeat do no exceed 5Hz and that the associated RMS motion amplitude of both cornea and retina does not exceed 1.5μm (sub-pixel in our case). A typical time for blood pressure-change-induced axial motion is about 1s. On the other hand, the time between each M-scan in our study is about 3ms (~256/90e3). As such, we can infer that recordings are not highly influenced by axial motion of the eye. In addition, the low frequency filtering induced by windowing the *x-t* plot removes the residual low frequency component. Finally, because the top-surface displacement is in-fact an averaged and weighted displacement from the first 200μm below the surface of the cornea, small motion effects can be disregarded. Bulk motion in awake animal and human models would likely require faster scan times or surface correction algorithms to account for micrometer and millimeter scale movements.

Although OCT can provide micron-scale spatial resolution, reconstructed quantitative mechanical properties for bounded materials such as the cornea have significantly degraded spatial resolution.[19,22] Dense-scanning of propagating elastic waves creates high-resolution images of local group velocity,[23] but these images do not map anisotropic elastic moduli at the same resolution. In reconstructing quantitative elastic moduli using guided waves, the resolution is determined by (i) the elastic wavelength of the generated mechanical wave[22] and (ii) the corneal thickness.[19] For AµT excitation, the theoretical elastic wavelengths generated can yield spatially resolved elastic moduli maps with mm-scale resolution. Demonstration of local anisotropic elastic moduli remains an area of future focus.

An appropriate mechanical model is required to reconstruct elastic moduli from experimental wave propagation data. Mechanical anisotropy has previously been confirmed in both ex vivo porcine and human cornea, where the in-plane tensile (Young's) modulus, $E$, is on average multiple orders of magnitude larger than the out-of-plane shear modulus, $G$. The NITI model used in this work introduced two independent shear moduli, separating in-plane from out-of-plane moduli. It helps to explain the order(s) of magnitude difference in corneal stiffness estimates extracted from shear- and tensile-based mechanical measurements. The results of this study show that AµT- OCE can be used to quantify anisotropic mechanical properties in living corneal tissue. As reported above, anisotropic mechanical moduli in rabbit corneas measured via OCE under both in vivo and ex vivo conditions were in close agreement with the values measured via destructive ex vivo testing.

The propagation speed of a vertically polarized bulk or surface shear wave in the cornea plane of a NITI medium is determined by the modulus, $G$, only. As such, the shear modulus, $\mu$, cannot be determined in a semi-infinite or bulk material. However, guided elastic wave propagation in the cornea plane also creates a weak dependence on $\mu$ through boundary conditions.[8] As shown here and in our recent work,[7,8] a change in $\mu$ results in a small change in the low-kHz region of the $A_0$-mode dispersion. Thus, a deviation of the NITI model from the actual cornea may further worsen the goodness of fit and result in a large confidence interval for reconstructed modulus $\mu$. A recent study[28], for example, suggests that the wave dispersion can also be affected by corneal prestress induced by IOP and its in-plane boundary conditions, which are not taken into account in our study. . Although the reconstruction of $G$ seems to be quite stable to small inaccuracies in the model used here, this model can be further refined for better quantification of $\mu$. Therefore, the reconstruction of $\mu$ from experimental data is less accurate compared to the reconstruction of $G$. Under these circumstances, it is important to evaluate whether the measured elastic waveform can be appropriately described using the NITI model. To determine whether $\mu$ can be accurately determined assuming a NITI model, a standardized method to determine data inclusion and model appropriateness is very important. In this work we developed a method to quantify data quality based on what we refer to as the goodness of fit (between measured and theoretical dispersion curves). Using a statistical analysis of experimental data, we have shown that the reconstruction of $G$ and $\mu$ can be considered correct for any fit when $g_{\text{NITI}}$ is above 0.87, and 0.92, respectively. Details on the determination of exclusion parameters can be found in Supplementary Methods.

Another important note is that this work assumed tensile isotropy of elastic moduli in the cornea, i.e. $E_T = E_L = 3\mu$. Most likely, this assumption requires further refinement of the NITI model, particularly in the definition of the proportionality coefficient between $E_T$ and $\mu$. Since corneal shear anisotropy is extremely strong, we expect this relationship to be closer to $E_T = 2\mu$. Measuring the exact relationship on a large population of cornea samples will be a part of our future studies. Here, however, we prefer to keep the relationship $E = E_T = E_L = 3\mu$ until it is accurately measured.

The cornea is also unique in that the fibers are pre-stressed by the intraocular pressure. Both $G$ and $E$ increased with intraocular pressure for in vivo and ex vivo measurements. Additionally, the stress-strain curves in Supplementary Materials show that tensile Young's Modulus $E$ is strain dependent. Clearly, the pre-strain condition will play a role in measured corneal stiffness. As an aside, it has also been shown that increased stress plays a role in the propagation behavior of elastic waves, even for linear elastic materials.[24,29] Different techniques have shown the possibility of disentangling the effect of IOP induced pre-stress from the change of stiffness for linearly isotropic solids.[28-30] Decoupling non-linear mechanical responses from strain-induced changes in wave speed for anisotropic materials remains an area of future interest.

In this work we also assumed that both $E$ and $G$ are real moduli, i.e. the loss modulus related to tissue viscosity is assumed negligibly small. As can be seen in the rheometry plots shown in Supplementary Methods, $G$ has a non-zero loss modulus and the frequency response is consistent with a measurable viscosity. Methods to estimate corneal viscosity (such as by tracking elastic wave energy) are currently under investigation. However, we notice here that cornea viscosity mostly affects the higher frequency range of the dispersion relationship for the $A_0$ mode,[25,26] and thus has very little effect on values estimated from the real part of moduli. In other words, the cornea loss modulus is not measured in this study, but introducing complex elastic moduli into the NITI model would produce small changes in the reconstructed elastic (real) part of moduli.

The in vivo measurement of intraocular pressure was performed using a contact Tono-Pen. Consistent with previously reported results,[20,21] the mean tonometry-measured IOP generally underestimated the actual IOP at high pressures, with very large individual variability between samples. One limitation in the present study was that only 5 samples were used to create the correction factor. Due to large individual variations, a more accurate correction factor would require a much larger population sample size. Because contact tonometers (such as the Tono-Pen) use a simple model to estimate internal pressure based on displacement of the cornea, the final value it measures is inherently linked to not only IOP, but also to cornea biomechanics. Independent measurements of corneal elastic moduli may help to develop better mechanical models that can potentially lead to more accurate estimates of IOP.

In comparing measured moduli between in vivo and ex vivo studies and between OCE measurement and mechanical tests, we note that sample boundary conditions were quite different for the different methods. In vivo scans were performed with the cornea intact, where the whole globe was held in place via ocular muscles and under normal IOP. Ex vivo scans attempted to match in vivo conditions; however, whole globes were no longer vascularized and ocular muscles no longer attached. Even more significant, rheometry and tensile testing required that the cornea was excised from the sclera and physically altered. Such alterations change the macro-structure and can release tension within lamellae. Additionally, the loading direction was different in ex vivo samples. Thus, identical values of moduli extracted with the different methods used in this study was not expected. Still, the results were in relatively close agreement.

In addition, shear rheometry and OCE methods determine the out-of-plane shear modulus, $G$, at different frequencies. Whereas rheometry operates in the Hz-range, OCE operates in the multiple-kHz regime. As can be seen in the individual plots shown in Supplementary Methods, the apparent modulus, $G$, increases with increasing frequency, suggesting that the OCE results reported in Figure 5 would be higher than the expected shear modulus of the cornea under lower frequency shear strain, as well as under typical biological shear strain rates (due to eye rubbing, for example). However, the results obtained with all methods were still in relatively close agreement, which supports the NITI model for reconstruction of cornea moduli from OCE data.

We also note a decrease in fit quality with increasing IOP (Supplementary Methods). Because increasing IOP has been shown to introduce an additional degree of anisotropy,[27] rabbit cornea may not be accurately described by the NITI model at high IOP (above 15 mmHg), and an orthotropic model (and, therefore, a larger number of elastic moduli) may be required.

**Conclusions**

Humans and rabbits share many common genetic features and by examining the physiology, anatomy and mechanical structures of the living rabbit, scientists can gain valuable insights into human function. In this study, the anisotropic shear moduli of rabbit corneas were measured in living animals and validated using multiple ex vivo measurement techniques on the same set of samples. The results of this work suggest that accurate in vivo measurements of human corneal moduli can be made using A$\mu$T-OCE, an important step toward supporting clinical adoption.

**Acknowledgements**

The authors wish to thank Nicholas Reyes at UW Animal Facility Operations and Gary Fye at UW Veterinary Services for veterinarian assistance.

**Data Availability**

The authors declare that all data from this study are available within the Article and its Supplementary Information. Raw data for the individual measurements are available upon reasonable request.

**Non-contact acoustic micro-tapping optical coherence elastography for quantification of corneal anisotropic elasticity: in vivo rabbit study**

Mitchell A. Kirby[1], Gabriel Regnault[1*], Ivan Pelivanov[1], Matthew O'Donnell[1], Ruikang K. Wang[1,3], Tueng T. Shen[2,3]

[1]Department of Bioengineering, University of Washington, Seattle, Washington 98105, USA
[2]School of Medicine, University of Washington, Seattle, Washington 98195, USA
[3]Department of Ophthalmology, University of Washington, Seattle, Washington 98104, USA
*Correspondence: Gabriel Regnault, Department of Bioengineering, University of Washington, Seattle, Washington 98105, USA Email: gregnaul@uw.edu

**Supplemental Methods**

**1. Quality of fit and generation of uncertainty intervals**

While $g_{\text{NITI}}$ (Eq. 5, see Methods) provides an estimate for how well the theoretical $A_0$ mode matches experimental data, it alone does not provide confidence intervals on the output moduli $G$ and $\mu$. Due to the maximization approach for weighted fitting based on the energy in the 2-D Fourier transform use in this study, residual errors are not computed, making traditional confidence interval methods difficult to apply. For example, a low value of $g_{\text{NITI}}$ would suggest that the $A_0$ dispersion curve calculated from the NITI model poorly described the actual dispersion measured within the sample. In such case, modulus estimates should have increased uncertainty.

To estimate uncertainty, $G$ and $\mu$ were varied independently around the values near a maximum $g_{\text{NITI}}$ and $\Phi(\mu, G)$ was recorded. For each combination of $G$ and $\mu$, Eq. 4 (Methods) was used to calculate $\Phi(\mu, G)$, which was then used to determine:

$$\psi(\mu, G) = \frac{\Phi(\mu, G)}{\Phi_{\max}}, \tag{S.1}$$

where the $\psi$ function represented goodness of fit values normalized to the maximum energy for a range of values. A representative example of $\psi(\mu, G)$ can be seen in Figure S1. Here (*x-t* plot shown in Figure S1a), the iterative routine converged on a best-fit $A_0$ mode (Figure S1b) where $g_{\mathrm{NITI}} = 0.98$ when $G = 20.5$ kPa and $\mu = 4$ MPa. In Figure S1c, $\psi$ is shown for $G \in [10 - 40]$ kPa and $\mu \in [1 - 50]$ MPa. It highlights the high sensitivity of $\psi$ to a change in $G$ while it remains relatively stable while $\mu$ varies. In Figure S1d, $\psi$ is shown when $\mu = 4$ MPa and $G$ was varied from 10 kPa- 40 kPa. The corresponding $A_0$ dispersion curves (constitutive equation found in Ref[2]) as $G$ varied can be seen in Figure S1f. As described previously, the high-frequency threshold of the $A_0$ dispersion curve is largely determined by $G$. In Figure S1e, $\psi$ is shown when $G = 20.5$ kPa and $\mu$ swept across a 1 MPa - 50 MPa range. Again, the corresponding $A_0$ dispersion curves can be seen in Figure S1g. As suggested previously, the $A_0$ mode is not as sensitive to changes in $\mu$ for the degree of shear anisotropy ($\mu/G$) expected in the cornea. As such, $\psi$ is less sensitive to changes in $\mu$ and produces higher relative uncertainty (see also Figure S1c). This routine provided a range for both $G$ and $\mu$ values indicating the degree to which the iterative solution converged on a single value.

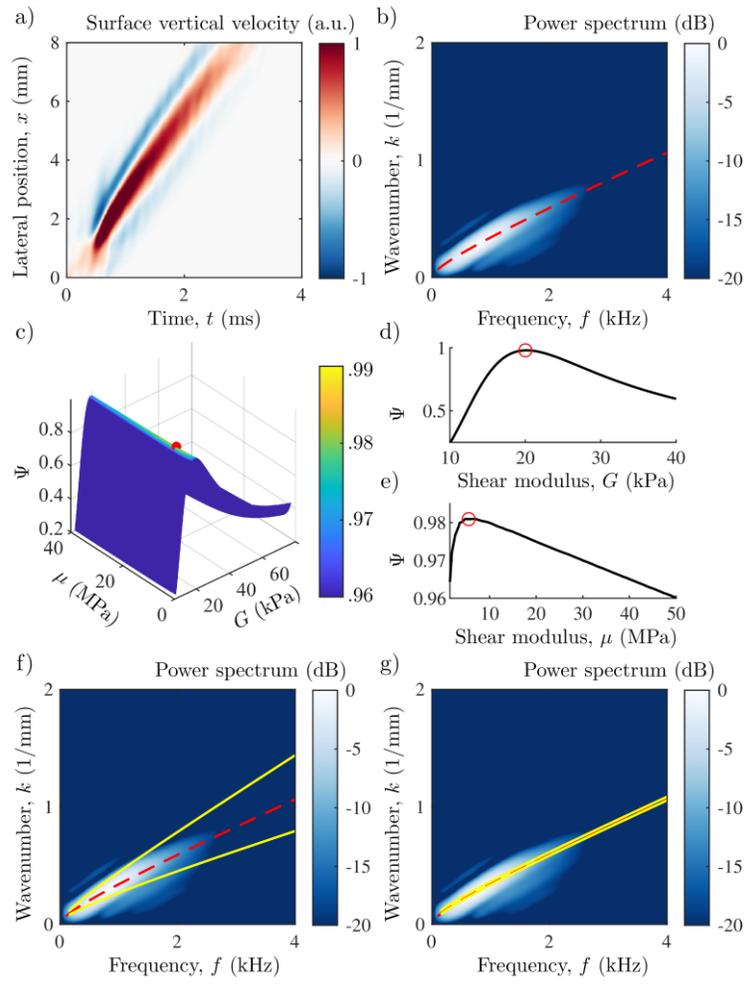

**Figure S1.** a) Space-time (*x-t*) plot of surface vibrations measured via OCT in a normal cornea sample. b) Best-fit solution to the dispersion equation in *k-f* space (based on a unique combination of elastic moduli, $\mu$ and $G$, displayed in red) on top of experimentally obtained A$_0$ mode for the corresponding cornea; c) Surface plot of $\psi$ for $G \in [10 - 40]$ kPa and $\mu \in [1 - 50]$ MPa. d) $\psi$ for $\mu = 4$ MPa, where $G$ was swept across a range from 10 kPa to 40 kPa. The red dot shows the highest value of $\psi$ e) $\psi$ for $G = 20.5$ kPa, where $\mu$ was swept across a range of 1 MPa - 50 MPa. Yellow lines plotted on top of measured energy in the *k-f* domain are A$_0$ dispersion curves corresponding to the range

f) $G$ = 10 kPa and $G$ = 40 kPa, with $\mu = 4$ MPa and g) $\mu$= 1 MPa and $\mu$= 50 MPa, with $G = 20.5$ kPa.

Due to the shape of $\psi$, the uncertainty in the fit produced uneven error bars. Note that the absolute value of $g_{\text{NITI}}$ provides an estimate for model error, where $g_{\text{NITI}} = 1$ for a NITI material would have very small uncertainty intervals. As $g_{\text{NITI}}$ is reduced, the shape of $\psi$ widens for both moduli and model uncertainty increases.

Because n=5 repeat scans were taken, five independent measurements produced corresponding values for $G$ and $\mu$ and their respective uncertainty ranges. The uncertainty ranges for each cornea (at each IOP) were finally calculated by taking the square-root of the means of the upper and lower limits, divided by the number of scans (n=5).

## 2. Exclusion criteria

In cases where the NITI model does not describe measured wavefields properly (due to poor excitation, misalignment, corneal structure abnormalities, etc.), the iterative fitting routine will arrive at a modulus value that has little to no physical meaning. To determine cases where the measured wavefields are accurately described by the NITI model, a 'cut-off' criterion in the goodness of fit for both $G$ and $\mu$ was determined. Any scan with a goodness of fit below the cut-off value was omitted from analysis.

To determine the relationship between $g_{\text{NITI}}$ and model error in rabbit corneas, an iterative fit was performed and $g_{\text{NITI}}$ calculated for all OCE scans (at all pressures), providing $g_{\text{NITI}}$/moduli histograms (Figures S2a-b). Each section of the histograms corresponds to a 0.01 range in $g_{NITI}$, 5 kPa for $G$, and 2 MPa for $\mu$. These histograms illustrate the repartition of moduli obtained from the fits and show, as expected, a stable repartition of moduli in the high goodness of fit range.

In order to determine critical $g_{NITI}$ values, histograms were integrated in the $g_{NITI}$ direction. For every modulus range, the number of occurrences was counted and the mean $g_{NITI}$ computed. These results are shown in Figures S2c-d. The modulus clearly increases (associated with faster wave speeds) as $g_{NITI}$ decreases progressively before approaching a point where this behavior breaks, reaching a plateau. The plateau is associated with nearly random statistics for reconstructed moduli, corresponding to the range of $g_{NITI}$ below which the fitting procedure used for moduli reconstruction is inaccurate. The final cutoff value was computed as the average $g_{NITI}$ over the plateau range. Table S1 presents the cutoff goodness of fit ($g_{NITI}$) criteria determined by the method detailed above for both $G$ and $\mu$

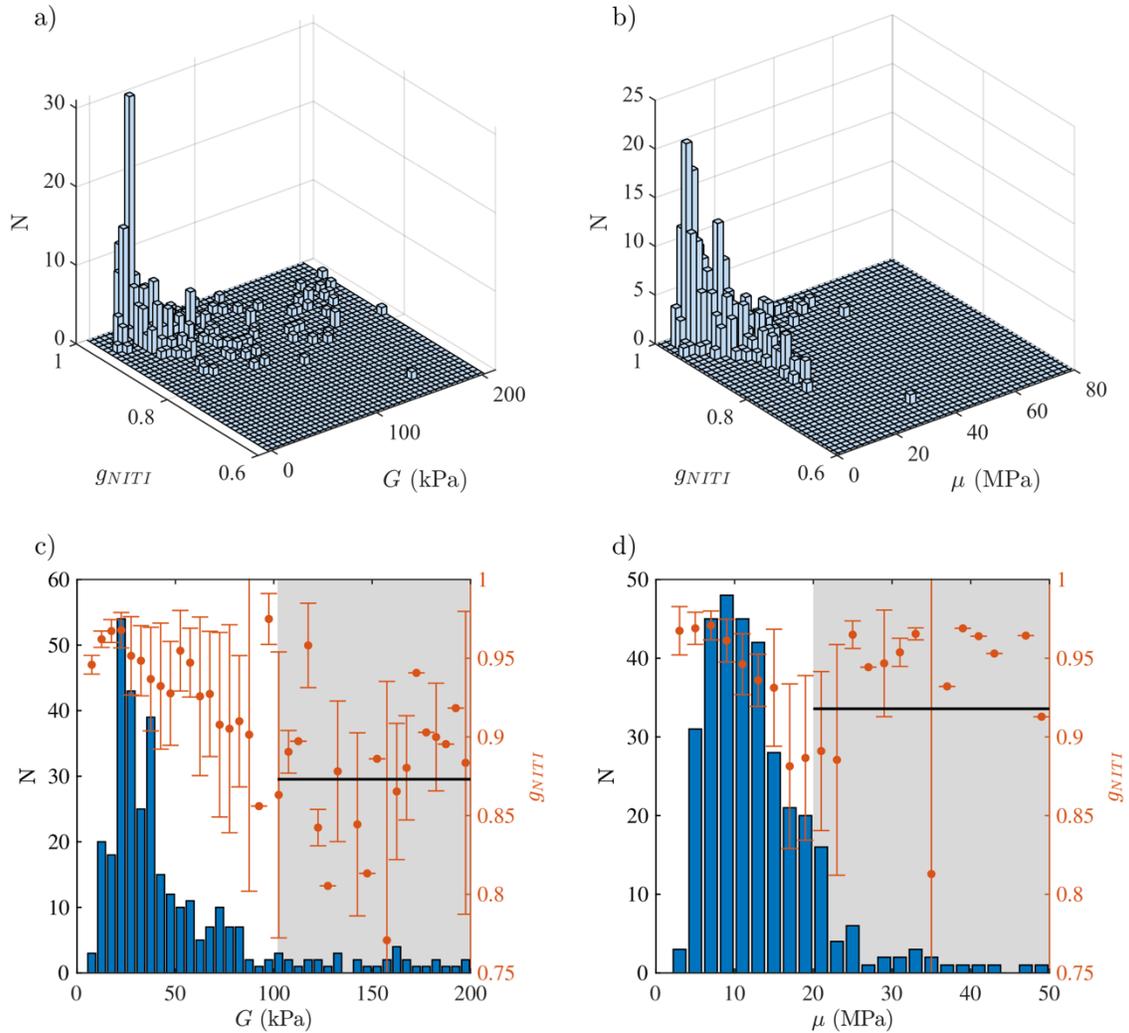

**Figure S2.** Procedure to determine the criteria of exclusion for $G$ and $\mu$. a), b) 2D histograms illustrating the distribution of fitted moduli, respectively for $G$ and $\mu$, and the goodness of fit, $g_{\text{NITI}}$, metric. c), d) 1D histogram illustrating the distribution of data as a function of the fitted moduli, respectively for $G$ and $\mu$. The right axis represents the averaged $g_{\text{NITI}}$ in the considered modulus range and the error bars represent the standard deviation in the measured goodness of fit. The black line and gray shaded area indicate the range over which the cutoff $g_{\text{NITI}}$ was calculated. Exact values for the cutoff $g_{\text{NITI}}$ are given in Table S1).

**Table S1:** Cut off values of goodness of fit, $g_{NITI}$

| Modulus | Cutoff value ($g_{NITI}$) |
|---------|---------------------------|
| $G$     | 0.87                      |
| $\mu$   | 0.92                      |

## 3. Mechanical testing results for individual cornea samples

Each cornea was first tested with OCE in vivo, where at least 10 non-contact OCE scans were performed on each eye. Once in vivo measurements were performed, the pressure was measured with a Tono-Pen. Following in vivo measurements, euthanasia was performed and whole globe corneas were harvested. Whole globes were placed in a mold containing a damp sterile cotton pad to stabilize samples and mimic in vivo boundary conditions. A 20-gauge needle connected to a bath filled with BSS was inserted through the temporal wall of the sclera to apply a controlled internal hydrostatic pressure (IOP) ranging from 1 mmHg to 23 mmHg. Each sample was scanned at room temperature and imaging took no longer than 1 hour per sample. After AµT-OCE, corneal buttons were cut and rheometry measurements were performed. Each corneal button then was sectioned into strips approximately 6 mm wide along the nasotemporal axis of the cornea and subject to tensile testing up to 10% strain. The elastic moduli quantified from all tests, for all corneas, are shown below (Figure S3- Figure S11). All ex vivo data were acquired within 6 hours of animal euthanasia. The corresponding $g_{NITI}$ for all OCE measurements are also included in Figure S12- Figure S20.

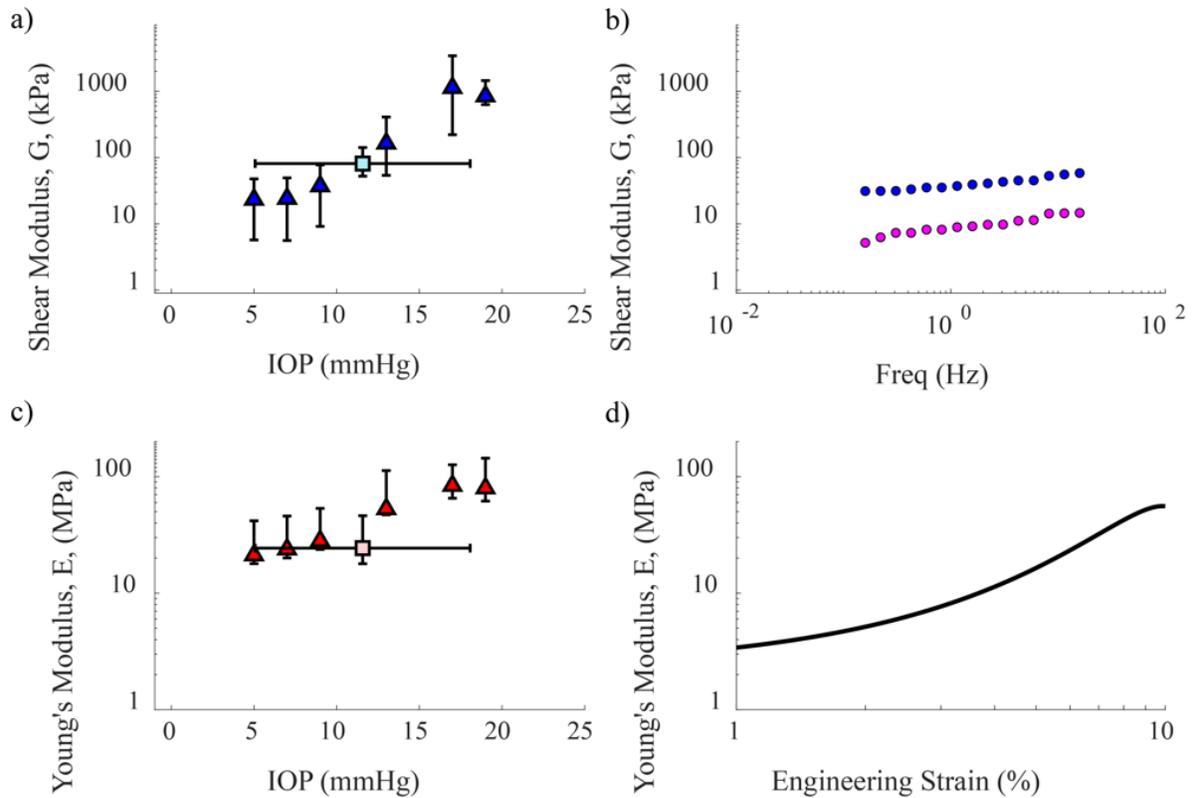

**Figure S3.** Measurement of anisotropic elastic moduli in Cornea #1. a) Out-of-plane shear modulus, $G$, measured with AµT-OCE in vivo (square) and ex vivo at controlled pressure (triangle). The vertical error bars correspond with uncertainty intervals and the horizontal error bar corresponds with in vivo IOP uncertainty. b) Out-of-plane shear modulus, $G$, measured with parallel plate rheometry. Blue corresponds with storage modulus and pink with loss modulus. c) In-plane Young's modulus, $E$, measured with AµT-OCE in vivo (square) and ex vivo at controlled pressure (triangle). The vertical error bars correspond with uncertainty intervals and the horizontal error bar corresponds with in vivo IOP uncertainty. d) Strain-dependent Young's moduli, $E$, measured via extension testing up to 10% strain, or where visible tissue damage occurred.

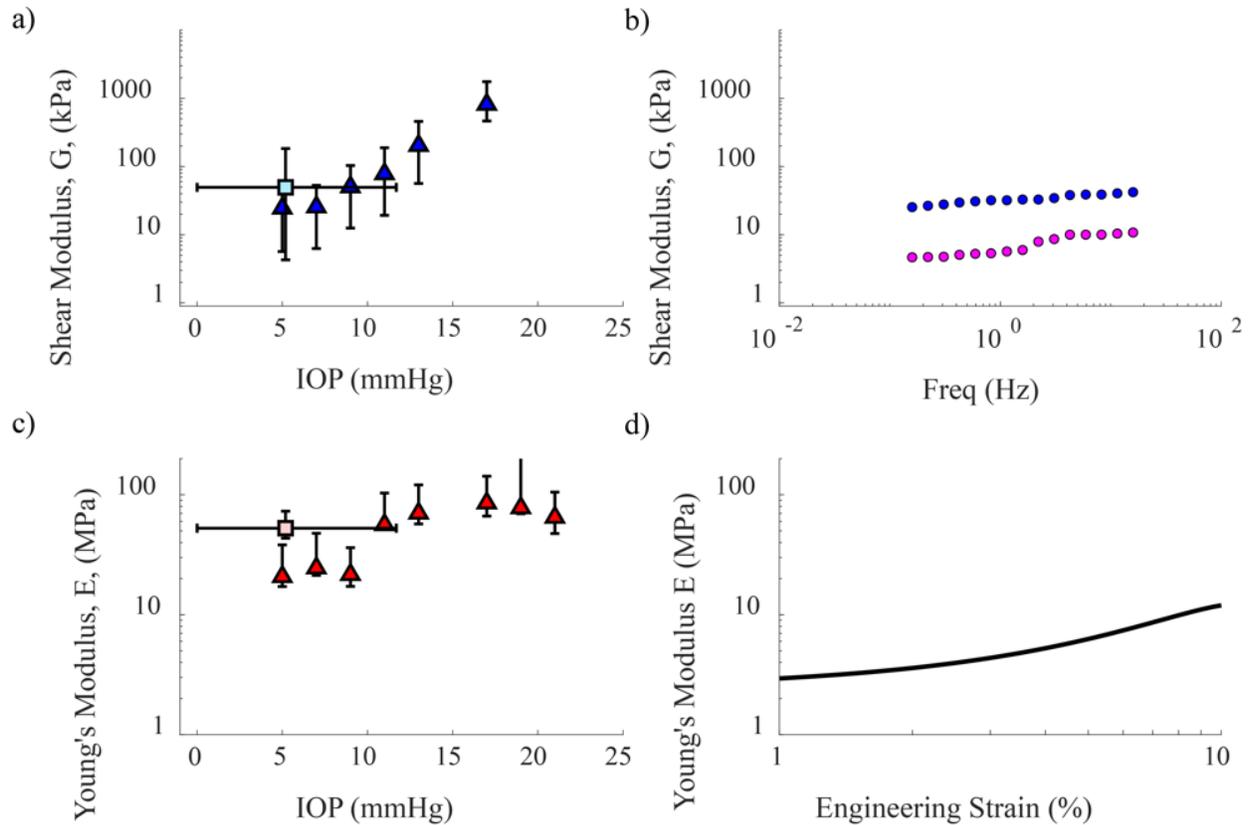

**Figure S4.** Measurement of anisotropic elastic moduli in Cornea #2. a) Out-of-plane shear modulus, $G$, measured with AµT-OCE in vivo (square) and ex vivo at controlled pressure (triangle). The vertical error bars correspond with uncertainty intervals and the horizontal error bar corresponds with in vivo IOP uncertainty. b) Out-of-plane shear modulus, $G$, measured with parallel plate rheometry. Blue corresponds with storage modulus and pink with loss modulus. c) In-plane Young's modulus, $E$, measured with AµT-OCE in vivo (square) and ex vivo at controlled pressure (triangle). The vertical error bars correspond with uncertainty intervals and the horizontal error bar corresponds with in vivo IOP uncertainty. d) Strain-dependent Young's moduli, $E$, measured via extension testing up to 10% strain, or where visible tissue damage occurred.

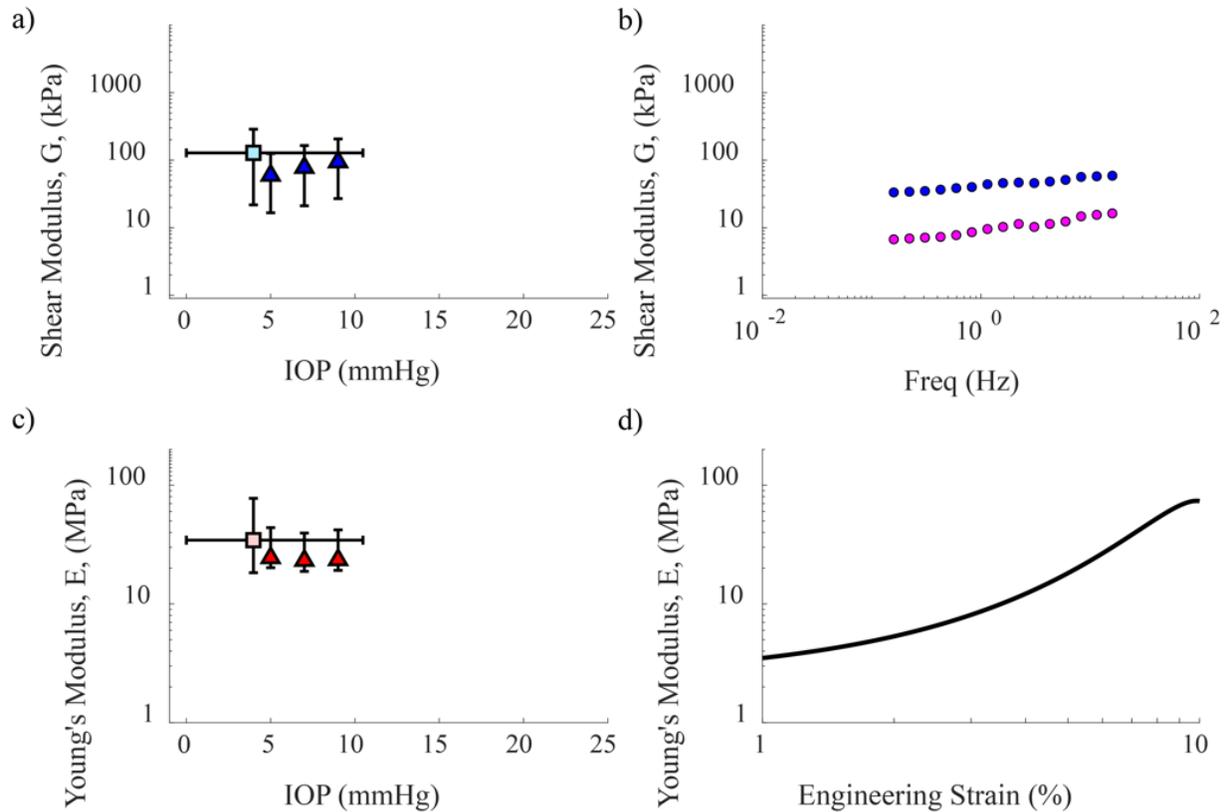

**Figure S5.** Measurement of anisotropic elastic moduli in Cornea #3. a) Out-of-plane shear modulus, $G$, measured with AµT-OCE in vivo (square) and ex vivo at controlled pressure (triangle). The vertical error bars correspond with uncertainty intervals and the horizontal error bar corresponds with in vivo IOP uncertainty. b) Out-of-plane shear modulus, $G$, measured with parallel plate rheometry. Blue corresponds with storage modulus and pink with loss modulus. c) In-plane Young's modulus, $E$, measured with AµT-OCE in vivo (square) and ex vivo at controlled pressure (triangle). The vertical error bars correspond with uncertainty intervals and the horizontal error bar corresponds with in vivo IOP uncertainty. d) Strain-dependent Young's moduli, $E$, measured via extension testing up to 10% strain, or where visible tissue damage occurred.

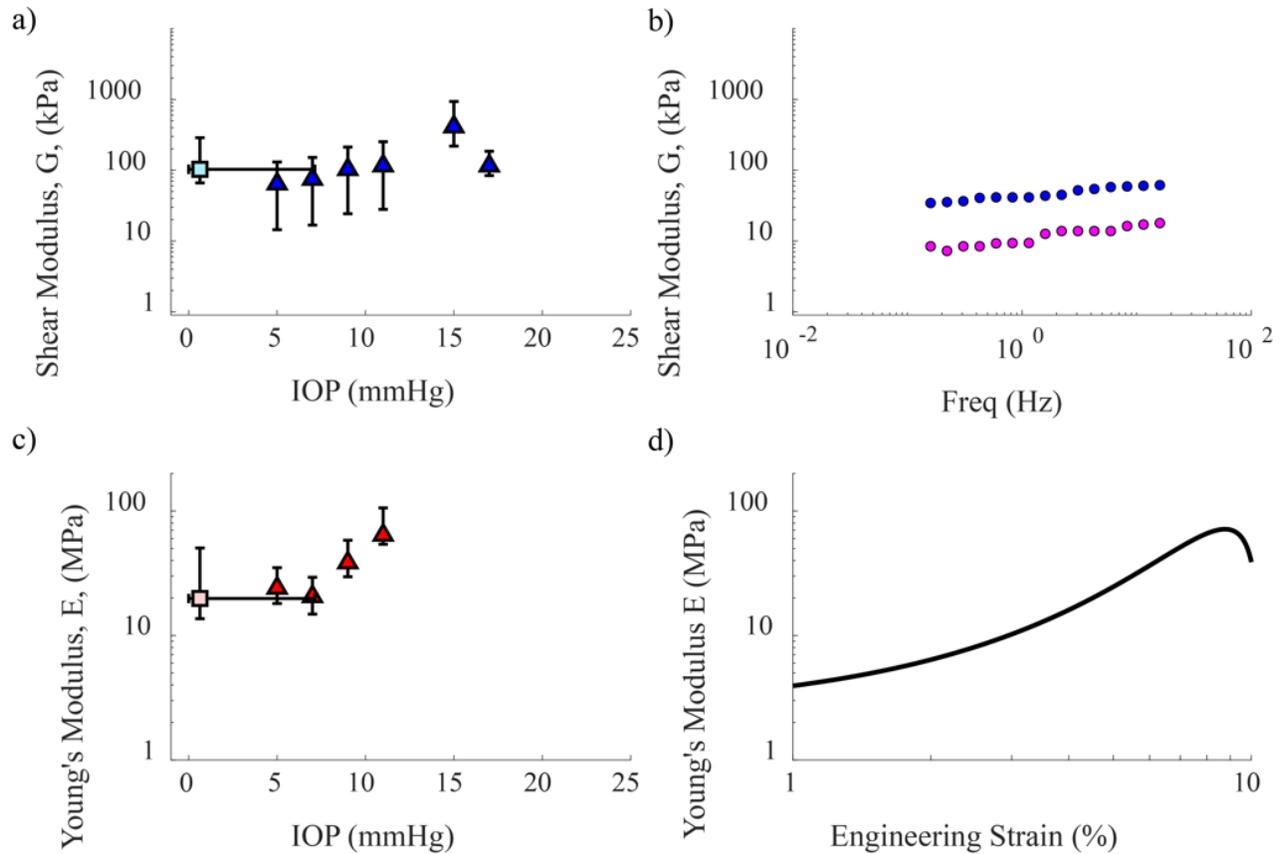

**Figure S6.** Measurement of anisotropic elastic moduli in Cornea #4. a) Out-of-plane shear modulus, $G$, measured with AµT-OCE in vivo (square) and ex vivo at controlled pressure (triangle). The vertical error bars correspond with uncertainty intervals and the horizontal error bar corresponds with in vivo IOP uncertainty. b) Out-of-plane shear modulus, $G$, measured with parallel plate rheometry. Blue corresponds with storage modulus and pink with loss modulus. c) In-plane Young's modulus, $E$, measured with AµT-OCE in vivo (square) and ex vivo at controlled pressure (triangle). The vertical error bars correspond with uncertainty intervals and the horizontal error bar corresponds with in vivo IOP uncertainty. d) Strain-dependent Young's moduli, $E$, measured via extension testing up to 10% strain, or where visible tissue damage occurred.

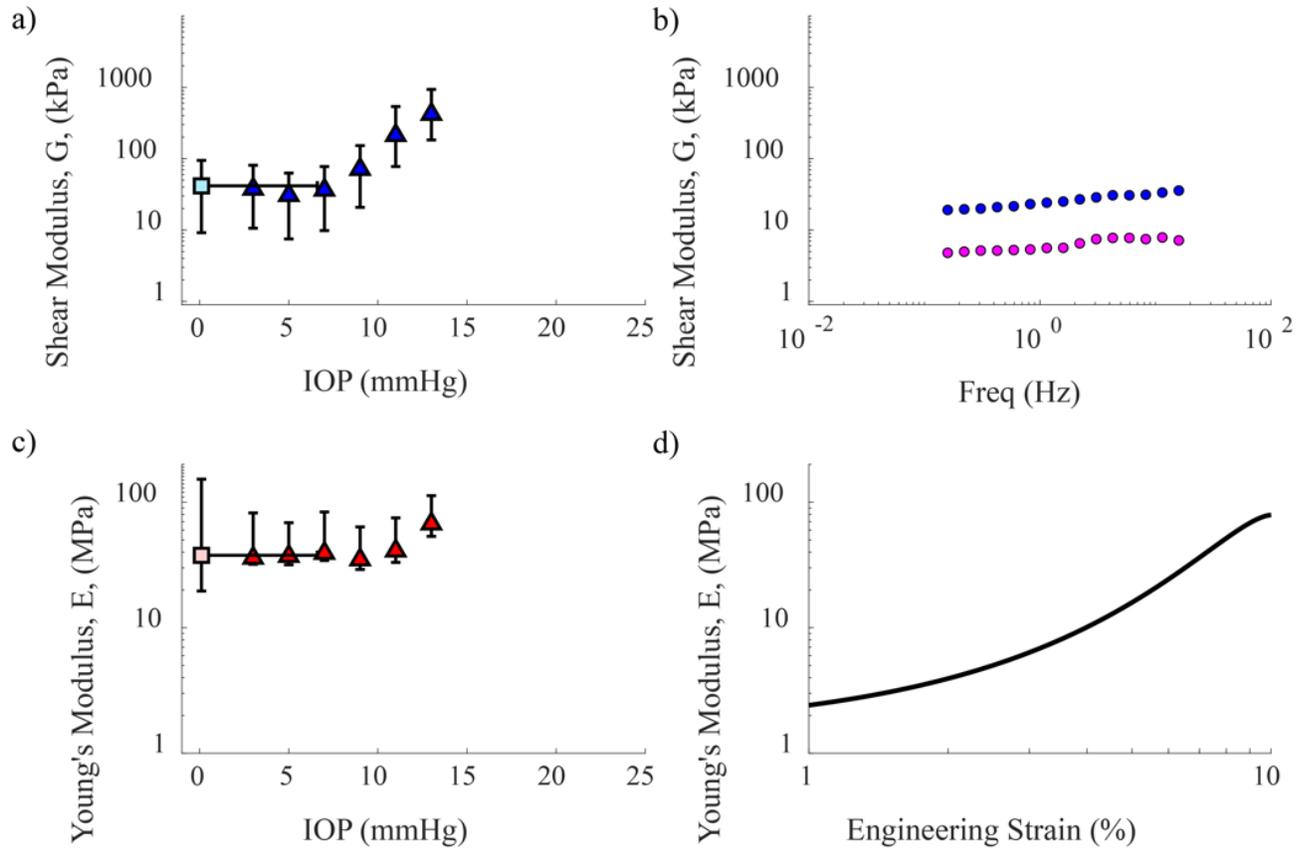

**Figure S7.** Measurement of anisotropic elastic moduli in Cornea #5. a) Out-of-plane shear modulus, $G$, measured with AµT-OCE in vivo (square) and ex vivo at controlled pressure (triangle). The vertical error bars correspond with uncertainty intervals and the horizontal error bar corresponds with in vivo IOP uncertainty. b) Out-of-plane shear modulus, $G$, measured with parallel plate rheometry. Blue corresponds with storage modulus and pink with loss modulus. c) In-plane Young's modulus, $E$, measured with AµT-OCE in vivo (square) and ex vivo at controlled pressure (triangle). The vertical error bars correspond with uncertainty intervals and the horizontal error bar corresponds with in vivo IOP uncertainty. d) Strain-dependent Young's moduli, $E$, measured via extension testing up to 10% strain, or where visible tissue damage occurred.

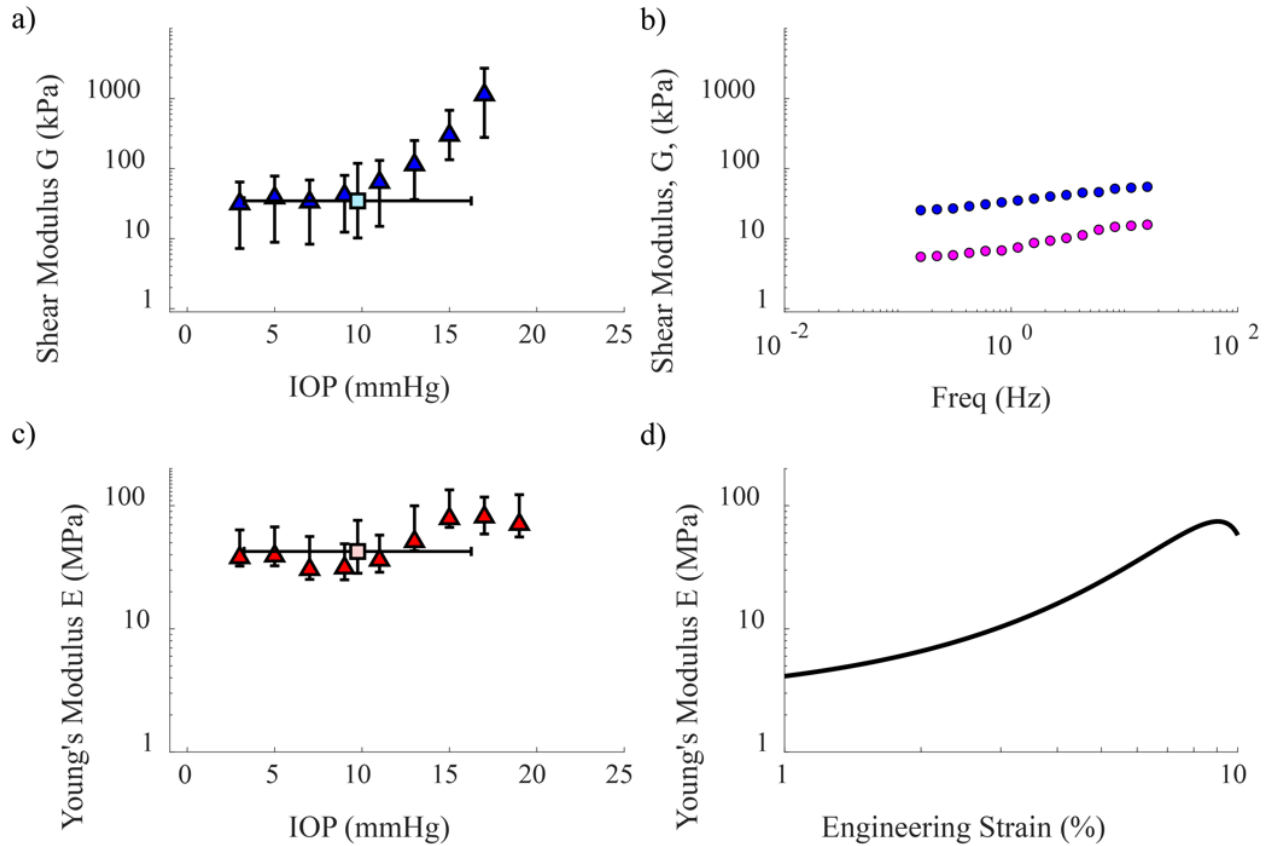

**Figure S8.** Measurement of anisotropic elastic moduli in Cornea #6. a) Out-of-plane shear modulus, $G$, measured with AμT-OCE in vivo (square) and ex vivo at controlled pressure (triangle). The vertical error bars correspond with uncertainty intervals and the horizontal error bar corresponds with in vivo IOP uncertainty. b) Out-of-plane shear modulus, $G$, measured with parallel plate rheometry. Blue corresponds with storage modulus and pink with loss modulus. c) In-plane Young's modulus, $E$, measured with AμT-OCE in vivo (square) and ex vivo at controlled pressure (triangle). The vertical error bars correspond with uncertainty intervals and the horizontal error bar corresponds with in vivo IOP uncertainty. d) Strain-dependent Young's moduli, $E$, measured via extension testing up to 10% strain, or where visible tissue damage occurred.

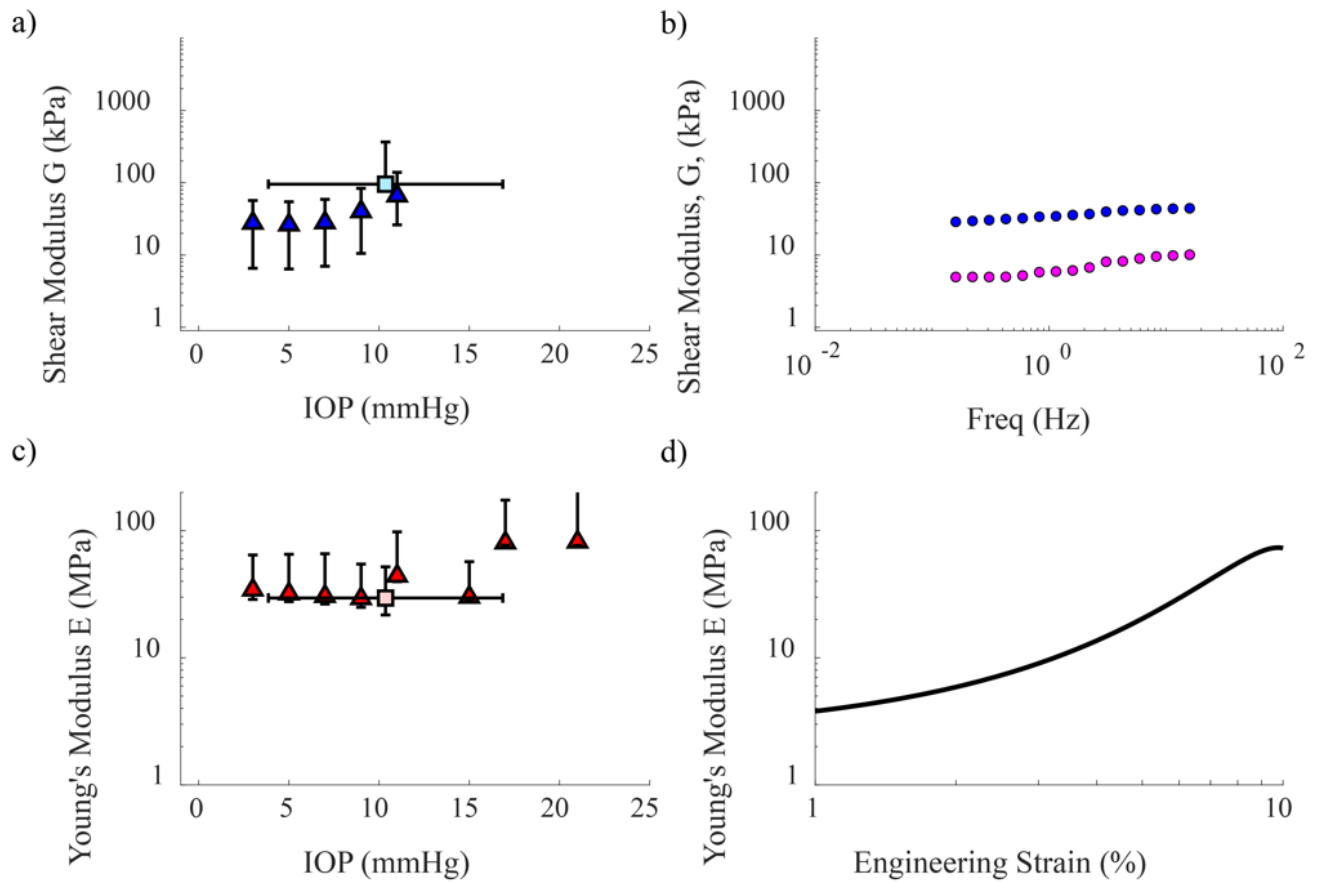

**Figure S9.** Measurement of anisotropic elastic moduli in Cornea #7. a) Out-of-plane shear modulus, $G$, measured with AµT-OCE in vivo (square) and ex vivo at controlled pressure (triangle). The vertical error bars correspond with uncertainty intervals and the horizontal error bar corresponds with in vivo IOP uncertainty. b) Out-of-plane shear modulus, $G$, measured with parallel plate rheometry. Blue corresponds with storage modulus and pink with loss modulus. c) In-plane Young's modulus, $E$, measured with AµT-OCE in vivo (square) and ex vivo at controlled pressure (triangle). The vertical error bars correspond with uncertainty intervals and the horizontal error bar corresponds with in vivo IOP uncertainty. d) Strain-dependent Young's moduli, $E$, measured via extension testing up to 10% strain, or where visible tissue damage occurred.

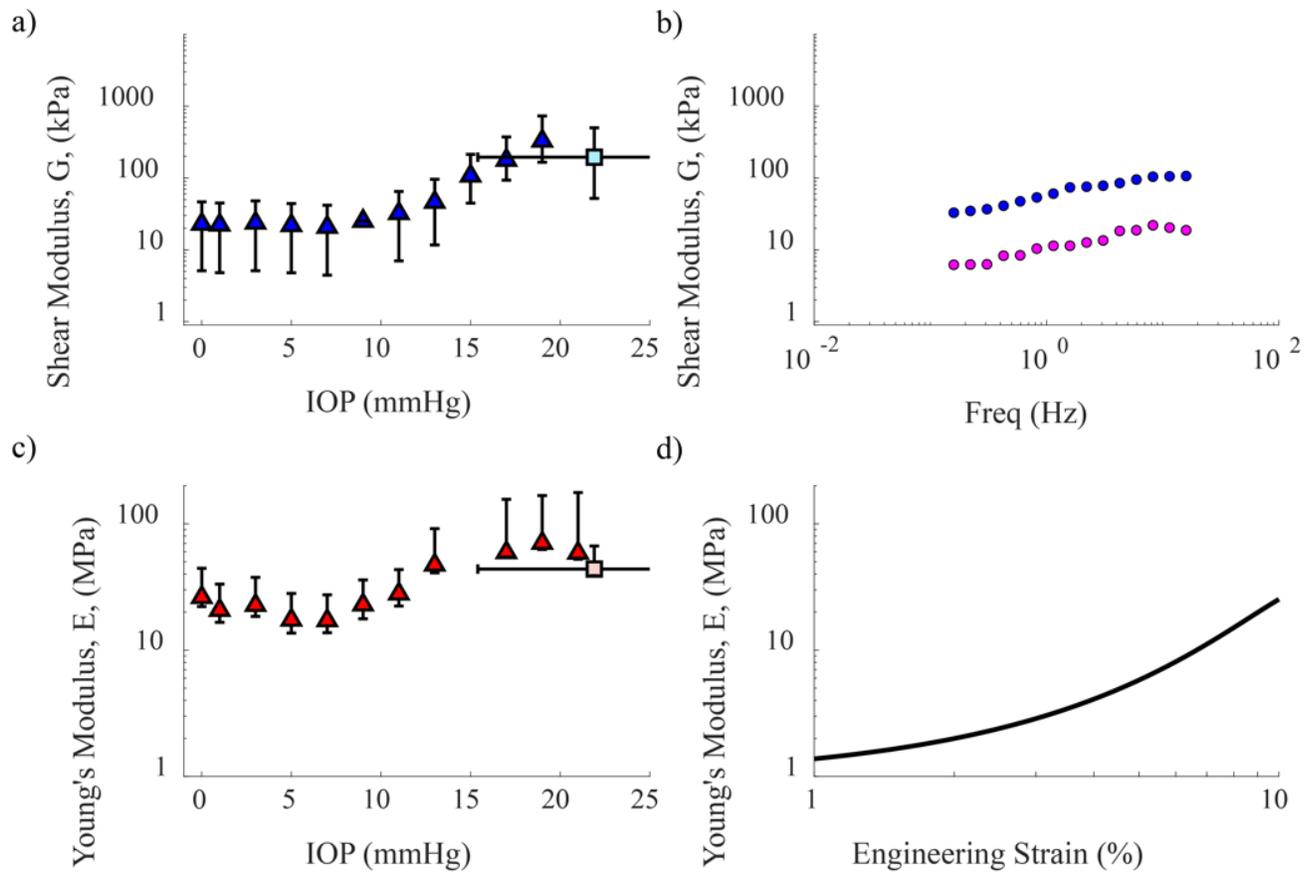

**Figure S10.** Measurement of anisotropic elastic moduli in Cornea #8. a) Out-of-plane shear modulus, $G$, measured with AµT-OCE in vivo (square) and ex vivo at controlled pressure (triangle). The vertical error bars correspond with uncertainty intervals and the horizontal error bar corresponds with in vivo IOP uncertainty. b) Out-of-plane shear modulus, $G$, measured with parallel plate rheometry. Blue corresponds with storage modulus and pink with loss modulus. c) In-plane Young's modulus, $E$, measured with AµT-OCE in vivo (square) and ex vivo at controlled pressure (triangle). The vertical error bars correspond with uncertainty intervals and the horizontal error bar corresponds with in vivo IOP uncertainty. d) Strain-dependent Young's moduli, $E$, measured via extension testing up to 10% strain, or where visible tissue damage occurred.

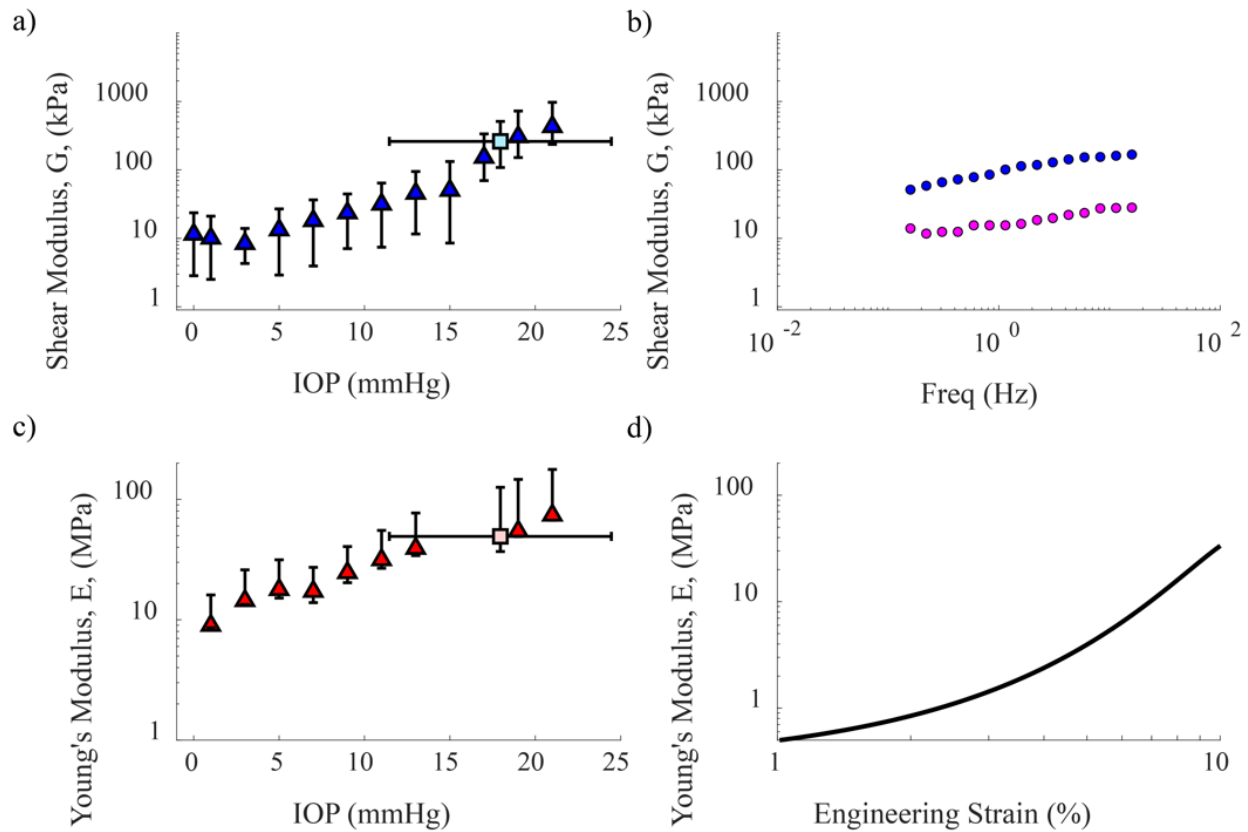

**Figure S11.** Measurement of anisotropic elastic moduli in Cornea #9. a) Out-of-plane shear modulus, $G$, measured with AμT-OCE in vivo (square) and ex vivo at controlled pressure (triangle). The vertical error bars correspond with uncertainty intervals and the horizontal error bar corresponds with in vivo IOP uncertainty. b) Out-of-plane shear modulus, $G$, measured with parallel plate rheometry. Blue corresponds with storage modulus and pink with loss modulus. c) In-plane Young's modulus, $E$, measured with AμT-OCE in vivo (square) and ex vivo at controlled pressure (triangle). The vertical error bars correspond with uncertainty intervals and the horizontal error bar corresponds with in vivo IOP uncertainty. d) Strain-dependent Young's moduli, $E$, measured via extension testing up to 10% strain, or where visible tissue damage occurred.

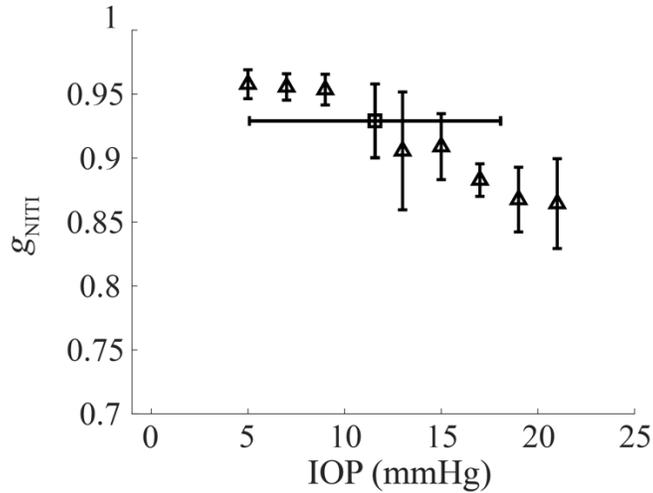

**Figure S12.** Mean goodness of fit ($g_{NITI}$) for all OCE scans in Cornea #1. The triangles correspond with ex vivo measurements at controlled IOP and error bars correspond with standard deviation across 5 repeat scans. The square corresponds with in vivo measurements and vertical error bars associate with the standard deviation of $g_{NITI}$ across at least 5 repeat scans. Horizontal error bar corresponds with in vivo IOP uncertainty.

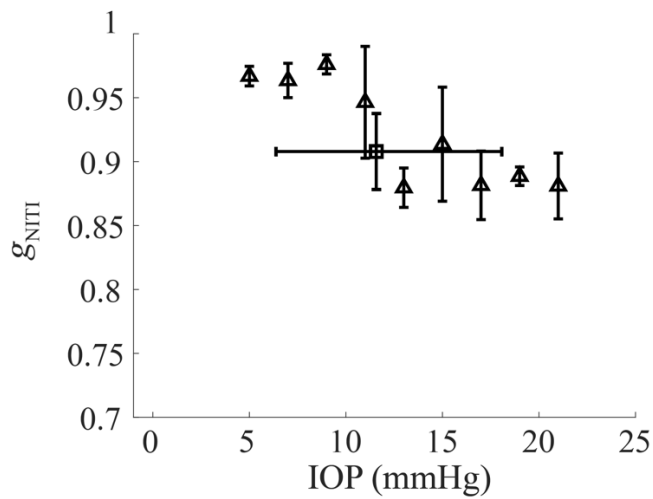

**Figure S13.** Mean goodness of fit ($g_{NITI}$) for all OCE scans in Cornea #2. The triangles correspond with ex vivo measurements at controlled IOP and error bars correspond with standard deviation across 5 repeat scans. The square corresponds with in vivo measurements and vertical error bars associate with the standard deviation of $g_{NITI}$ across at least 5 repeat scans. Horizontal error bar corresponds with in vivo IOP uncertainty.

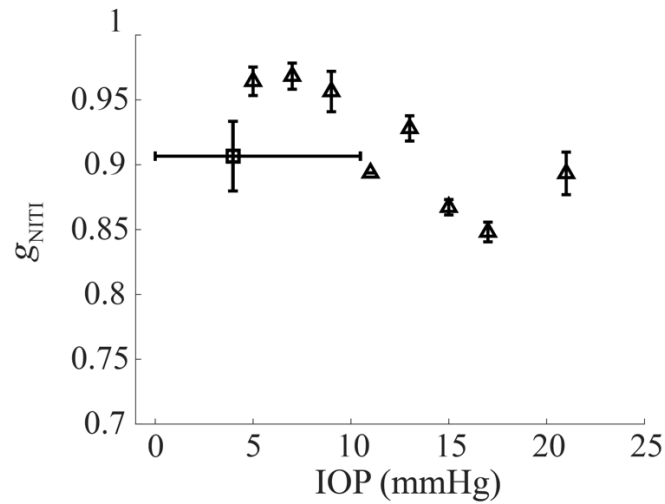

**Figure S14.** Mean goodness of fit ($g_{NITI}$) for all OCE scans in Cornea #3. The triangles correspond with ex vivo measurements at controlled IOP and error bars correspond with standard deviation across 5 repeat scans. The square corresponds with in vivo measurements and vertical error bars associate with the standard deviation of $g_{NITI}$ across at least 5 repeat scans. Horizontal error bar corresponds with in vivo IOP uncertainty.

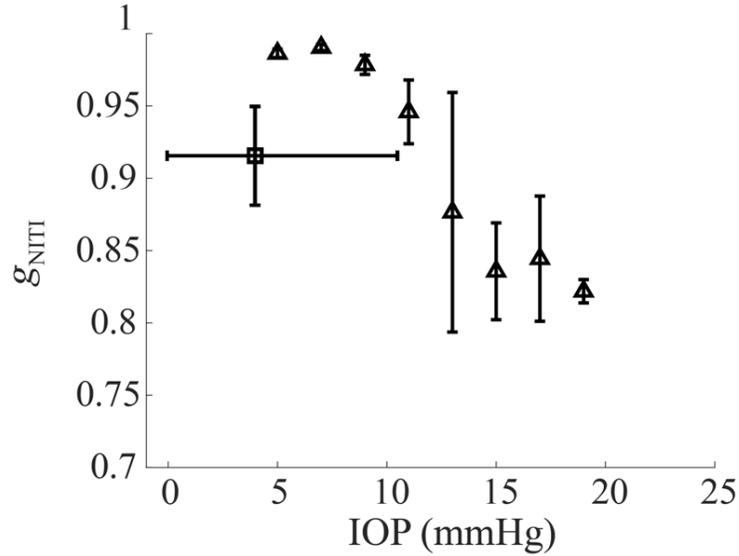

**Figure S15.** Mean goodness of fit ($g_{NITI}$) for all OCE scans in Cornea #4. The triangles correspond with ex vivo measurements at controlled IOP and error bars correspond with standard deviation across 5 repeat scans. The square corresponds with in vivo measurements and vertical error bars associate with the standard deviation of $g_{NITI}$ across at least 5 repeat scans. Horizontal error bar corresponds with in vivo IOP uncertainty.

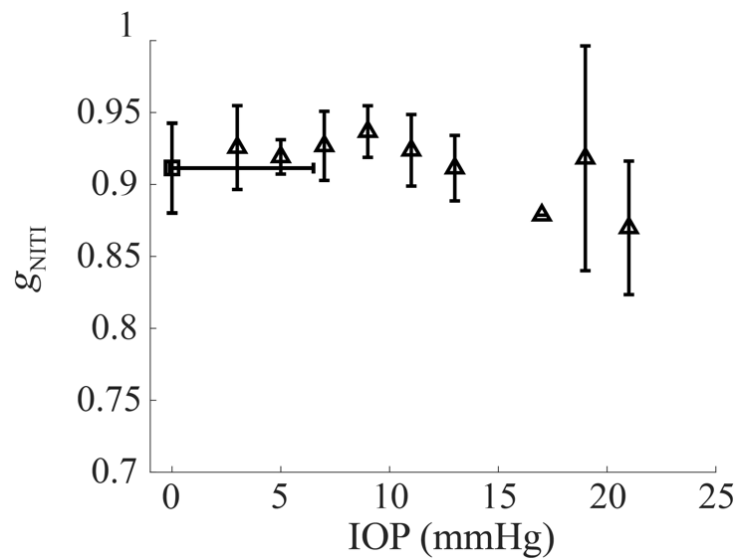

**Figure S16.** Mean goodness of fit ($g_{NITI}$) for all OCE scans in Cornea #5. The triangles correspond with ex vivo measurements at controlled IOP and error bars correspond with standard deviation across 5 repeat scans. The square corresponds with in vivo measurements and vertical error bars associate with the standard deviation of $g_{NITI}$ across at least 5 repeat scans. Horizontal error bar corresponds with in vivo IOP uncertainty.

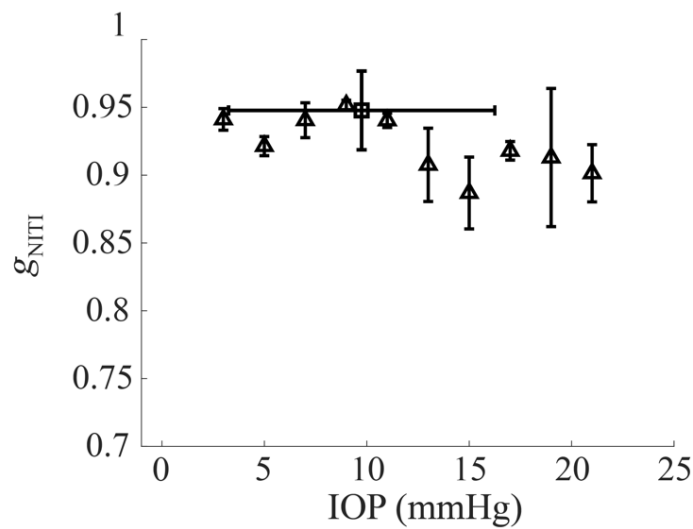

**Figure S17.** Mean goodness of fit ($g_{NITI}$) for all OCE scans in Cornea #6. The triangles correspond with ex vivo measurements at controlled IOP and error bars correspond with standard deviation across 5 repeat scans. The square corresponds with in vivo measurements and vertical error bars associate with the standard deviation of $g_{NITI}$ across at least 5 repeat scans. Horizontal error bar corresponds with in vivo IOP uncertainty.

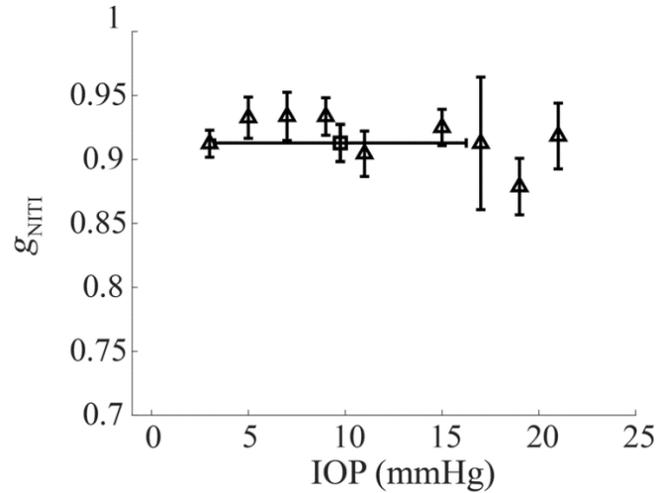

**Figure S18.** Mean goodness of fit ($g_{NITI}$) for all OCE scans in Cornea #7. The triangles correspond with ex vivo measurements at controlled IOP and error bars correspond with standard deviation across 5 repeat scans. The square corresponds with in vivo measurements and vertical error bars associate with the standard deviation of $g_{NITI}$ across at least 5 repeat scans. Horizontal error bar corresponds with in vivo IOP uncertainty.

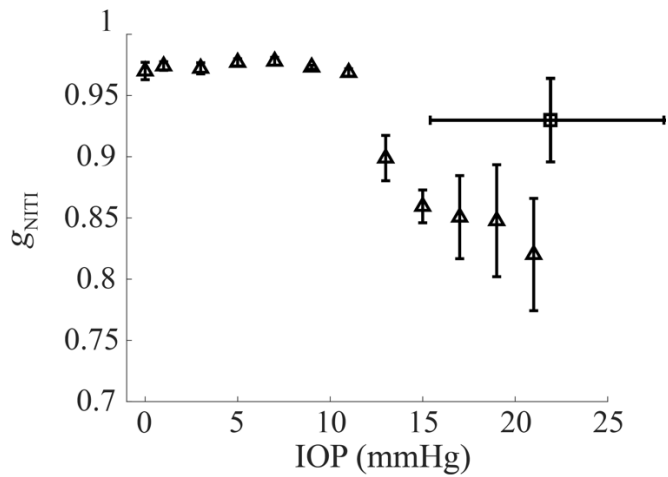

**Figure S19.** Mean goodness of fit ($g_{NITI}$) for all OCE scans in Cornea #8. The triangles correspond with ex vivo measurements at controlled IOP and error bars correspond with standard deviation across 5 repeat scans. The square corresponds with in vivo measurements and vertical error bars associate with the standard deviation of $g_{NITI}$ across at least 5 repeat scans. Horizontal error bar corresponds with in vivo IOP uncertainty.

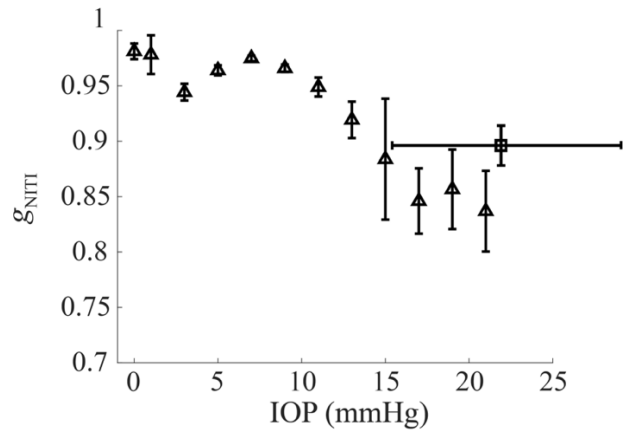

**Figure S20.** Mean goodness of fit ($g_{NITI}$) for all OCE scans in Cornea #9. The triangles correspond with ex vivo measurements at controlled IOP and error bars correspond with standard deviation across 5 repeat scans. The square corresponds with in vivo measurements and vertical error bars associate with the standard deviation of $g_{NITI}$ across at least 5 repeat scans. Horizontal error bar corresponds with in vivo IOP uncertainty.

**Supplemental References**